\documentclass[journal]{IEEEtran}
\usepackage[utf8]{inputenc}
\usepackage[T1]{fontenc}
\usepackage{cite}
\usepackage{graphicx}
\usepackage[caption=false,font=footnotesize]{subfig}
\usepackage{amsmath}
\usepackage{algorithm}
\usepackage[noend]{algpseudocode}
\usepackage{array}
\usepackage{booktabs}
\usepackage{breqn}
\usepackage{amssymb}
\usepackage{multirow}

\usepackage[table]{xcolor} 
\usepackage{xcolor}
\usepackage{dblfloatfix}
\usepackage{url}
\usepackage{xurl}
\usepackage[acronyms,nonumberlist,nopostdot,nomain,nogroupskip,acronymlists={hidden}]{glossaries}
\newacronym{ofdm}{OFDM}{Orthogonal Frequency-Division Multiplexing}
\newacronym{5g}{5G}{Fifth Generation}
\newacronym{fft}{FFT}{Fast Fourier Transform}
\newacronym{ifft}{IFFT}{Inverse Fast Fourier Transform}
\newacronym{isi}{ISI}{Intersymbol Interference}
\newacronym{ue}{UE}{User Equipment}
\newacronym{gnb}{gNB}{Next Generation Node B}
\newacronym{ta}{TA}{Timing Advance}
\newacronym{uav}{UAV}{Unmanned Aerial Vehicle}
\newacronym{srs}{SRS}{Sounding Reference Signal}
\newacronym{dmrs}{DMRS}{Demodulation Reference Signal}
\newacronym{ul}{UL}{Uplink}
\newacronym{rb}{RB}{Resource Block}
\newacronym{zc}{ZC}{Zadoff-Chu}
\newacronym{cazac}{CAZAC}{Constant Amplitude Zero Auto-Correlation}
\newacronym{cp}{CP}{Cyclic Prefix}
\newacronym{4g}{4G}{Fourth Generation}
\newacronym{6g}{6G}{Sixth Generation}
\newacronym{cfo}{CFO}{Carrier Frequency Offset}
\newacronym{bic}{BIC}{Bayesian Information Criterion}
\newacronym{papr}{PAPR}{peak-to-average power ratio}
\newacronym{snr}{SNR}{Signal-to-Noise Ratio}
\newacronym{pusch}{PUSCH}{Physical Uplink Shared Channel}
\newacronym{pucch}{PUCCH}{Physical Uplink Control Channel}
\newacronym{toa}{ToA}{Time-of-Arrival}
\newacronym{iid}{i.i.d.}{independent and identically distributed}
\newacronym{loa}{LOA}{Low-Altitude Economy}
\newacronym{los}{LoS}{Line-of-Sight}
\newacronym{nlos}{NLoS}{Non-Line-of-Sight}
\newacronym{nr}{NR}{New Radio}
\newacronym{aoa}{AoA}{Angle of Arrival}
\newacronym{prach}{PRACH}{Physical Random
 Access Channel}
 \newacronym{rss}{RSS}{Received Signal Strength}
 \newacronym{sdr}{SDR}{Software Defined Radio}
 \newacronym{tdoa}{TDoA}{Time Difference of Arrival}
 \newacronym{rssi}{RSSI}{Received Signal Strength Indicator}
 \newacronym{rf}{RF}{Radio Frequency}
 \newacronym{gcs}{GCS}{ground control station}
 \newacronym{udp}{UDP}{User Datagram Protocol}
 \newacronym{oai}{OAI}{OpenAirInterface}
 \newacronym{ssh}{SSH}{Secure Shell}
 \newacronym{ism}{ISM}{Industrial, Scientific, and Medical}
\newacronym{cpu}{CPU}{Central Processing Unit}
\newacronym{mavlink}{MAVLink}{Micro Air Vehicle Link}
\newacronym{srd}{SRD}{Short Range Device}
\newacronym{cots}{CotS}{Commercial off-the-Shelf}
\newacronym{wpbms}{WPBMS}{Weighted Power-Based Mean-Shift}
\newacronym{usrp}{USRP}{Universal Software Radio Peripheral}
\newacronym{dft}{DFT}{Discrete Fourier Transform}
\newacronym{idft}{IDFT}{Inverse Discrete Fourier Transform}
\newacronym{lawn}{LAWN}{Low-Altitude Wireless Networks}
\newacronym{uas}{UAS}{Unmanned Aereal System}
\newacronym{rrc}{RRC}{Radio Resource Control}

\begin{document}
%
\title{Multi-UE Identification and Localization in LAWN via an Autonomous Non-Serving UAV}
%
%
%

\author{Niccolò Paglierani~\IEEEmembership{Student Member,~IEEE}, Francesco Linsalata~\IEEEmembership{Member,~IEEE},  \\ Vineeth Teeda~\IEEEmembership{Member,~IEEE}, Davide Scazzoli~\IEEEmembership{Member,~IEEE}, Maurizio Magarini,~\IEEEmembership{Member,~IEEE} \thanks{The authors are with the Dipartimento di Elettronica, Informazione e Bioingegneria, Politecnico di Milano, 20133 Milan, Italy (e-mail name.surname@polimi.it). The authors thank the Polus Tech team for their support on the research subject and Dr. Christian Mazzucco for the deep discussion on the topics.}}

\maketitle

\begin{abstract}
This paper presents an autonomous sensing framework for identifying and localizing multiple User Equipments (UEs) in Fifth Generation (5G) cellular networks using a non-serving Unmanned Aerial Vehicle (UAV). In this context, autonomy refers to the UAV’s ability to perform synchronization, UE identification, localization, and mission execution entirely onboard, without external intervention or real-time network control. To this end, a complete onboard processing chain is developed, enabling synchronization, multi-UE identification, and localization directly from standard 3GPP-compliant uplink Sounding Reference Signals (SRS). Unlike conventional UAV-assisted approaches relying on serving nodes or infrastructure support, the proposed platform operates as a passive sensing UAV, requiring only minimal initial coordination with the network and no mission-time control-plane interaction.
The approach leverages the structured and periodic nature of SRS transmissions together with a tailored protocol configuration to ensure robust operation under realistic multi-UE interference conditions.
The system is designed to operate using narrowband SRS (1.4\,MHz), reducing UE power consumption and hardware complexity while enabling high multiplexing capability through combined use of cyclic shifts and frequency resources. Despite this low-complexity configuration, reliable synchronization and multi-UE identification are achieved even when multiple UEs share the same resources.
The UAV autonomously collects measurements along its trajectory and performs localization through a trajectory-based processing strategy. The approach is validated through extensive simulations and a full-scale experimental campaign, achieving localization errors below 8\,m in urban scenarios and below 3\,m in rural conditions, outperforming state-of-the-art Angle Of Arrival (AoA)- and Time Difference of Arrival (TDoA)-based methods by approximately 5--6\,m.
These results demonstrate the feasibility of infrastructure-independent sensing UAVs as a key enabler of Low-Altitude Wireless Networks (LAWN), supporting scalable and rapidly deployable situational awareness in emergency and connectivity-limited environments.
\end{abstract}

\begin{IEEEkeywords}
Low Altitude Economy (LAE), Unmanned Aerial Vehicle (UAV), Autonomous systems, Localization, Multi-UEs, 5G.
\end{IEEEkeywords}

%
\IEEEpeerreviewmaketitle

\section{Introduction}

\IEEEPARstart{T}{he low-altitude economy} is rapidly evolving thanks to the integration of \glspl{uas} with \gls{5g} and beyond terrestrial cellular networks, enabling aerial communication, sensing, and autonomous operations \cite{feng2025networked, kishk2020aerial}. In this context, \gls{lawn} extend cellular systems into the aerial domain, where \glspl{uav} can act as \glspl{ue}, serving nodes, or sensing platforms.

A relevant application of \gls{lawn} systems is the deployment of tactical wireless bubbles, i.e., temporary, mission-oriented, controllable network infrastructures established in emergency or disaster scenarios \cite{tact_bubb, telefonica_5g_nato_2026, microamp_tactical_bubbles_2025, albanese2021sardo, JSTSP_Disaster, moro2024enhancing, LinsalataOTFS}. In such contexts, beyond providing connectivity, a key requirement is to enable situational awareness through rapid, accurate \gls{ue} localization, even when terrestrial infrastructure is unavailable or impaired.

Conventional \gls{uav}-assisted localization approaches typically rely on aerial serving nodes, i.e., UAV-mounted base stations integrated within the cellular network, that establish communication links with ground \glspl{ue} \cite{albanese2021sardo, esrafilian2025first, mandloi2023q}. In these systems, the UAV has access to network synchronization and control-plane information, enabling network-assisted localization strategies. However, in tactical deployments, these platforms are primarily designed to maximize coverage and connectivity rather than optimize localization performance. Moreover, maintaining favorable propagation conditions (e.g., \gls{los}) for multiple spatially distributed \glspl{ue} is generally infeasible for a single serving UAV \cite{al2014optimal, s17020413, scazzoli2023experimental, JSAC_Paglierani}.

In light of these limitations, an alternative approach is to deploy UAVs as non-serving sensing nodes, operating independently from the communication infrastructure. In this configuration, the UAV passively receives \gls{ul} signals from connected \glspl{ue} and performs localization based solely on signal measurements, without participating in the communication process. This architectural decoupling enables greater flexibility in deployment, broader spatial coverage, and reduced hardware complexity, making it particularly suitable for rapid and scalable tactical operations.

These advantages, however, come at the cost of additional challenges. A non-serving UAV is not synchronized with the cellular network and has no access to scheduling information; therefore, it cannot predict when UE transmissions occur. As a result, synchronization, \gls{ue} identification, and localization must be performed directly from the received waveform, even in the presence of multi-\gls{ue} interference.

A fundamental gap in existing non-serving UAV localization studies lies in the mismatch between algorithmic assumptions and real cellular operation. Most prior works rely on idealized conditions, assuming perfect knowledge of \gls{ul} resource allocation or transmission timing \cite{petitjean2018fast, scazzoli2023experimental, 11162478}, which is not available to a passive sensing UAV.

In this work, we address this limitation by enabling multi-\gls{ue} localization directly from standard 3GPP-compliant \gls{ul} signals, without infrastructure modification and with only minimal initial coordination. The UAV operates as an autonomous sensing node, meaning that it performs synchronization, \gls{ue} identification, localization, and mission execution entirely onboard during flight, without external intervention or real-time network control. To this end, a complete onboard processing chain is developed, enabling synchronization, multi-\gls{ue} identification, and localization directly from standard 3GPP-compliant \gls{ul} \glspl{srs} \cite{TS38.211}, by exploiting their structured and periodic nature.

\textbf{Paper Contributions.}  
The main contributions of this paper are summarized as follows:
\begin{itemize}
 \item We introduce an autonomous low-complexity non-serving UAV sensing and localization framework for lightweight platforms.
The UAV operates as a passive receiver without network synchronization or control-plane access, executing the full sensing pipeline onboard,  without external intervention. The framework achieves sub-3\,m (rural) and sub-8\,m (urban) accuracy, outperforming \gls{aoa}- \cite{scazzoli2023experimental} and \gls{tdoa}-based \cite{esrafilian2025first} approaches by 5--6\,m despite using narrowband (1.4\,MHz) SRS.

 \item We develop a synchronization strategy based on SRS protocol design.
By optimizing the SRS configuration and subcarriers allocation, we enable reliable synchronization under multi-\gls{ue} interference. Analytical results show robust operation under worst-case multiplexing conditions over deployment scales up to $40$\,km.

 \item We reformulate multi-\gls{ue} identification as a frequency extraction problem.
Using SRS cyclic shifts and a Matching Pursuit approach, we achieve reliable identification with misidentification probability below $1\%$ in large-scale deployments for small \gls{ue} groups, and negligible in typical tactical scenarios. Scalability is ensured through multiplexing across multiple SRS subbands.

 \item We validate the proposed framework through simulations  and a full-scale outdoor campaign using a UAV equipped with an ADALM Pluto SDR \cite{plutoSDR_wiki}, a Pixhawk controller \cite{pixhawk_org}, and a Jetson Orin NX \cite{jetson_orin_nx}, demonstrating fully autonomous onboard operation.
\end{itemize}

\textbf{Paper Organization.} The remainder of the paper is structured as follows. Section~\ref{sec:related_work} reviews related works on experimental UAV-based localization. Section~\ref{sec:scenario_model} presents the system model, including UAV and \gls{ue} configurations and the \gls{srs} structure. In Sec.~\ref{sec:workflow}, we describe the proposed framework workflow, detailing synchronization, \gls{ue} identification, and measurement collection stages. Section~\ref{sec:results} reports numerical results evaluating synchronization, localization, and identification of multiple \glspl{ue} under different scenarios and parameters, while Sec.~\ref{sec:experiments} presents the testbed setup and experimental validation. Finally, Sec.~\ref{sec:conclusion} concludes the paper and outlines future work.

\textbf{Paper Notation.}  
Vectors and matrices are denoted by bold lowercase (\( \mathbf{x} \)) and uppercase (\( \mathbf{X} \)) letters, respectively, with transposition indicated by \( (\cdot)^T \). The notation \(\operatorname{diag}(\mathbf{x})\) represents a square matrix with the elements of \( \mathbf{x} \) on its main diagonal.  
The absolute value of a scalar \( x \) is \( |x| \), and the vector norm is \( \|\mathbf{x}\| \).  
The operators \( \lfloor x \rfloor \) and \( \lfloor x \rceil \) denote the floor and rounding to the nearest integer.  
A continuous-time signal \( x(t) \) sampled with period \( T_s \) is represented as \( x[n] = x(nT_s) \).  
Its \( N \)-point \gls{dft} is \( \mathrm{DFT}_N\{x[n]\} \), and the complex conjugate of \( x \) is \( x^* \).  
Sets are denoted by calligraphic uppercase letters (e.g., \( \mathcal{A} \)), with cardinality \( |\mathcal{A}| \).

\section{Related Works} 
\label{sec:related_work}

This section reviews prior research on UAV-assisted localization, focusing on experimental studies that provide practical benchmarks for real-world implementations. Many works are simulation-based and rely on idealized assumptions, while experimental studies reveal the challenges and limitations of operating UAVs in realistic cellular networks. The following discussion highlights these contributions and key constraints.

Research on UAV-assisted localization can be broadly divided into two categories. The first includes serving-UAV approaches, where the aerial platform acts as a base station or relay to provide both communication and positioning services, typically at the cost of increased system complexity, hardware requirements, and operational overhead. The second focuses on non-serving UAVs, which operate as passive receivers to estimate \gls{ue} positions without participating in the network, enabling more lightweight and cost-effective deployments, but relying on more limited information due to the lack of direct control-plane interaction and scheduling knowledge.

In serving-UAV systems, several experimental works have investigated UAV-assisted localization using time-based measurements. Esrafilian \textit{et al.} \cite{esrafilian2025first} presented one of the first real-world implementations of UAV-aided ToA localization in a 5G system using \gls{oai}, combining UAV motion with \gls{ul} SRS measurements to jointly estimate \gls{ue} and UAV positions via least-squares optimization. Similarly, Albanese \textit{et al.} \cite{albanese2021sardo} introduced SARDO, an autonomous UAV-based search-and-rescue system that localizes victims via their mobile phones without terrestrial infrastructure \cite{LinsalataOTFS}. The UAV carries a lightweight LTE base station and estimates positions using Demodulation Reference Signal (DMRS)-based pseudo-trilateration and neural networks. Both studies demonstrate the feasibility of UAV-assisted localization with real cellular signals but share limitations of \gls{toa} methods, including timing offsets from asynchronous UE clocks and the need for wide bandwidths, which may be impractical in emergencies.
The work in \cite{moro2024enhancing} explores the potential of an Integrated Sensing and Communication (ISAC) system to improve search-and-rescue operations, detailing methods for processing echoed signals from downlink transmissions to enhance target detection and classify possible victims by fusing Synthetic Aperture Radar (SAR) imagery with triangulation results from \gls{ul} signals \cite{JSTSP_SAR_loc, JSTSP_SAR}.

Other experimental approaches have explored non-serving \glspl{uav}. Scazzoli \textit{et al.} \cite{scazzoli2023experimental} demonstrated an \gls{aoa}-based method where a \gls{uav} receives \gls{prach} signals from a single \gls{ue}, achieving about $25\,$m accuracy. 
However, the system requires multiple antennas, and its \gls{aoa} estimates become unreliable during \gls{uav} rotations, as Doppler shifts distort the measured signal phase. \gls{prach} transmissions are also unpredictable, limiting their use for continuous non-serving UAV localization. 
Petitjean \textit{et al.} \cite{petitjean2018fast} proposed a \gls{rssi}-based method with a UAV equipped with a \gls{usrp} \gls{sdr} to estimate the bearing of a ground transmitter. While effective in \gls{los} conditions, it is sensitive to multipath and UAV altitude variations, and may converge to local \gls{rssi} maxima in complex environments. 
Masrur and Güvenç \cite{11162478} introduced a deep-learning framework trained on simulated data for 3D source localization, achieving around 18\,m average error in real-world trials. Table~\ref{tab:related_works} summarizes the main experimental studies in the literature, highlighting the achieved localization accuracy in rural scenarios, the considered deployment settings, and the degree of network involvement, including whether the UAV operates as a serving node and whether synchronization and \gls{ue} identification are available during operation.

\begin{table}[!t]
\captionsetup{justification=raggedright,singlelinecheck=false}
\caption{Experimental related works and their main limitations.}
\label{tab:related_works}
\renewcommand{\arraystretch}{0.8}
\setlength{\tabcolsep}{2.5pt}
\centering
\scriptsize
\rowcolors{2}{gray!15}{white}
\begin{tabular}{
p{0.06\linewidth}
p{0.15\linewidth}
p{0.09\linewidth}
p{0.08\linewidth}
p{0.12\linewidth}
p{0.13\linewidth}
>{\raggedright\arraybackslash}p{0.24\linewidth}
}
\toprule
\textbf{Ref.} & \textbf{Technique} & \textbf{Acc.} & \textbf{Serv.} & \textbf{Sync./ID} & \textbf{Scenario} & \textbf{Limitation} \\
\midrule
\cite{albanese2021sardo} & Trilateration & $\leq$50\,m & Y & Y & 300$\!\times\!$300\,m$^2$ & Infrastructure-dependent, low accuracy \\
\cite{moro2024enhancing} & RSSI \& SAR & $\leq$10\,m & Y & Y & 180$\!\times\!$180\,m$^2$ & Complex HW, high processing cost \\
\cite{esrafilian2025first} & TDoA & $\leq$15\,m & Y & Y & 70$\!\times\!$70\,m$^2$ & Tight synchronization, wide BW required \\
\cite{scazzoli2023experimental} & AoA Triang. & $\leq$50\,m & N & N & 100$\!\times\!$100\,m$^2$ & Antenna array calibration, rotation sensitivity \\
\cite{petitjean2018fast} & W-RSS & $\leq$10\,m & N & N & 25$\!\times\!$25\,m$^2$ & Sensitive to multipath, local maxima \\
\cite{11162478} & Deep Learn. & 20\,m & N & N & 170$\!\times\!$220\,m$^2$ &  Generalization \\
\bottomrule
\end{tabular}

\end{table}

Additional simulated studies highlight the benefit of trajectory optimization in \gls{uav} deployments. For example, \cite{s17020413}
analyzes how different UAV mobility patterns affect data collection, evaluating a time–coverage efficiency metric. They show that circular and square trajectories offer the best trade-off between area coverage and efficiency.
Similarly, \cite{christy2017optimum} evaluates several UAV flight paths for device discovery in post-disaster scenarios, showing that the most effective trajectory depends on the damage distribution.

These studies highlight the potential of UAV-assisted localization but also show practical limits, often due to partial autonomy, strict synchronization assumptions, and the need for complex or costly hardware.
In contrast, our work introduces a fully autonomous sensing UAV that performs synchronization, multi-\gls{ue} identification, localization, and mission planning entirely onboard using lightweight, low-cost hardware. 

To the best of the authors' knowledge, this is the first work that explicitly addresses onboard synchronization and \gls{ue} identification in a passive non-serving UAV setting, where no real-time control-plane information or prior knowledge of \gls{ue} transmissions is available.

\section{System Model} \label{sec:scenario_model}

In this section, we describe the scenario, the \gls{uav} deployment, the communication between ground \glspl{ue} and the network, and the properties of the \gls{ul} signals.

\subsection{Scenario and operational workflow}
\label{sec:Scenario}
\begin{figure}[!t]
    \centering
    \includegraphics[width=0.9\linewidth]{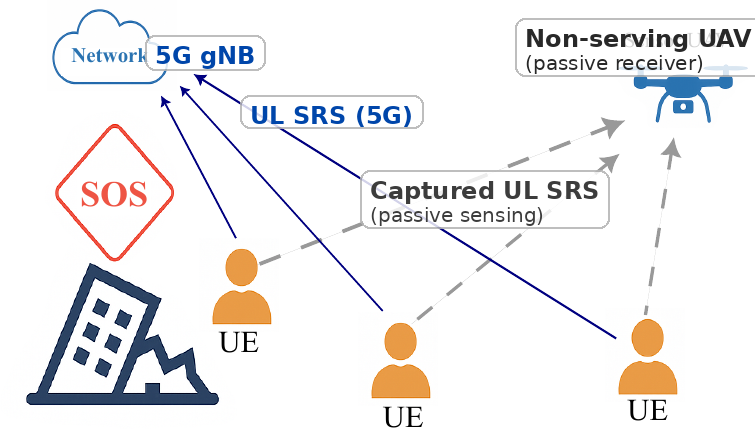}
  \caption{Operational scenario where a sensing UAV acts as a non-serving passive receiver to localize multiple UEs. The UAV passively captures UL SRS transmissions intended for the cellular network (gNB) without participating in the communication process, enabling infrastructure-independent sensing.}
    \label{fig:scenario}
\end{figure}

\begin{figure*}[!t]
    \centering
    \includegraphics[width=\textwidth]{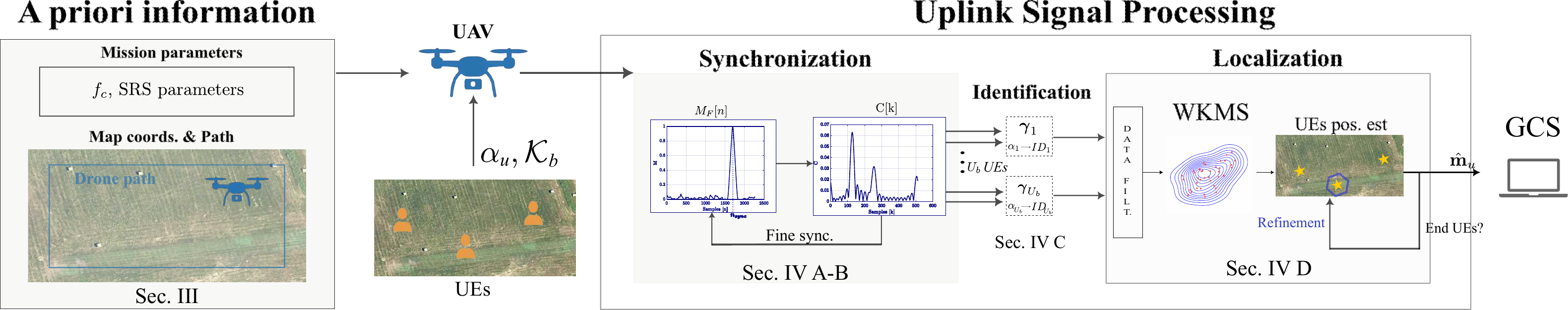}
\caption{Overview of the proposed UAV-based sensing framework. Starting from mission parameters and SRS configuration, the UAV acquires uplink SRS signals from multiple UEs along its trajectory. The onboard processing chain consists of: (i) synchronization, leveraging the SRS structure for coarse detection and fine time alignment; (ii) UE identification, exploiting frequency-domain processing and cyclic shift estimation to separate users; and (iii) localization, based on a weighted mean-shift algorithm that processes spatially distributed SRS-specific measurements collected along the UAV path. An initial coarse estimate is first obtained for all UEs simultaneously, followed by a refinement stage performed individually for each UE through autonomous trajectory updates. The resulting UE position estimates are finally transmitted to the GCS.}
    \label{fig:summary_scheme}
\end{figure*}

We consider the scenario in Fig.~\ref{fig:scenario}, where a sensing \gls{uav} is deployed in an emergency environment to localize $N_{\mathrm{UE}}$ \glspl{ue} transmitting 5G \gls{ul} signals toward a serving \gls{gnb}.
The position of each \gls{ue} $u $$\,\in\,$$ \mathcal{U}$, in the set of \glspl{ue} $\mathcal{U} $$\,=\,$$ \{ I_j \,| \,j = 1, \ldots, N_{\mathrm{UE}} \}$, is represented by the vector $\mathbf{m}_u $$\,=\,$$ [x_u, y_u, z_u]^T$, while the position of \gls{uav} evolves over time and is denoted, at the $r$th measurement time step, as $\mathbf{p}[r] $$\,=\,$$ [x^{\mathrm{UAV}}[r], y^{\mathrm{UAV}}[r], z^{\mathrm{UAV}}[r]]^T$.
The \gls{uav} therefore collects a set of $R$ measurements during its trajectory, corresponding to successive time instants along the flight path.
The \glspl{ue} are connected to the network through a \gls{gnb}, exchanging radio frames that also include the transmission of \gls{srs}. 
Figure~\ref{fig:summary_scheme} illustrates the overall mission workflow, which will be detailed in the following sections. 

A downlink telemetry connection is established with a \gls{gcs} allowing the network to track estimated \gls{ue} positions, identities, and \gls{uav} status parameters such as position and velocity. 
At the beginning of the mission, the network provides the \gls{uav} with the \gls{srs} configuration parameters (detailed in Sec.~\ref{subsec:SRS}) for all active \glspl{ue}, enabling \glspl{ue} localization and identification from the received signals. The \gls{uav} is also given an a priori area, defined by its center and extension, to guide its operation.
The UAV is equipped with \(N_{\text{ant}}\) antennas, collected in the set
\(\mathcal{A} = \{ A_{\ell} \mid \ell = 1, \dots, N_{\text{ant}} \}\), which are arranged to maximize spatial diversity. In the considered UAV platform (see Sec.~\ref{subsec:testbed config}), this is achieved by mounting the antennas on the landing legs, spaced approximately \(30\,\text{cm}\) apart and oriented at a \(90^\circ\) angle with respect to each other to mitigate polarization effects.
After the mission initialization phase, the workflow consists of three main stages using the received \glspl{srs}: \textit{(i)} coarse and fine time synchronization, \textit{(ii)} \gls{ue} identification and measurement extraction, and \textit{(iii)} localization of the identified \glspl{ue}.
\subsubsection*{Complexity analysis of the three processing stages} The synchronization step relies on a recursive update and is computationally negligible, with complexity $\mathcal{O}(1)$ per received sample, and no further re-synchronization is required thereafter.
The \gls{ue} identification, based on the Matching Pursuit algorithm \cite{wang2012generalized} and \gls{dft} operations, dominates the overall cost with complexity \(\mathcal{O}(L U_b \log L)\) per valid measurement, where \(L\) is half the symbol \gls{dft} length and \(U_b\) the number of \glspl{ue} in the \gls{srs} subband $\mathcal{K}_b$. 
Finally, the localization stage has complexity \(\mathcal{O}(U_b\, i_{\max}^{(\mathrm{loc})})\) per valid measurement, with \(i_{\max}^{(\mathrm{loc})}\) denoting the maximum number of iterations.

\subsection{Sounding Reference Signals (SRS)}
\label{subsec:SRS}

\glspl{srs} are pilot signals used in \gls{5g} mainly to evaluate the \gls{ul} channel quality and, starting from 3GPP Release 16, also for positioning purposes \cite{JSAC_Paglierani, TS38.211}. They are therefore highly configurable in both time and frequency domains. Unlike other \gls{ul} signals transmitted by different \glspl{ue}, whose time–frequency resources are dynamically assigned by the \gls{gnb} scheduler in an implementation-dependent manner, \glspl{srs} can be configured with fixed resources. For example, they may be set with a periodicity $T_{per}$ as short as a single \gls{ofdm} slot, and with an arbitrary subcarrier starting position $k_0$, occupying a minimum bandwidth $m_{SRS}$ of 4 \glspl{rb}. A smaller bandwidth offers several advantages in rescue scenarios, such as enhancing the \gls{snr} for the same transmitted power, reducing receiver complexity, improving the multiplexing capability of the network, and facilitating parallel search operations among multiple \glspl{uav}.

\begin{figure}[!t]
    \centering
    \includegraphics[width=0.9\linewidth]{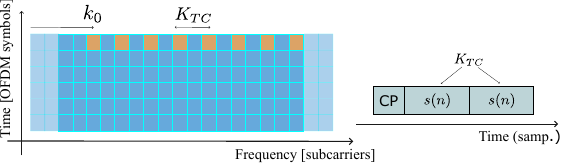}
    \caption{Illustration of the \gls{srs} mapping: (left) placement on the \gls{ofdm} grids with $K_{TC}=2$, and (right) the corresponding repeated structure in the time domain.}
    \label{fig:srs_grid}
\end{figure}

Another distinctive feature is that the SRS modulation sequence $Q[k]$ is mapped onto subcarriers spaced by multiples of $K_{TC}$, where $K_{TC}\in\{2,4,6,8,12\}$, leaving the remaining subcarriers unused. Throughout this paper, we consider the minimum spacing $K_{TC}=2$, as further motivated in Sec.~\ref{sec:Coarse time sync} and Sec.~\ref{sec: user identification}. The resulting sequence length is therefore $M_{\mathrm{SRS}}=12m_{\mathrm{SRS}}/K_{TC}$. An example of the resulting resource allocation is shown on the left of Fig.~\ref{fig:srs_grid}.

The transmitted SRS sequence is obtained by applying a phase increment to a base sequence generated from the cell identifier $q$, and can be written as \cite{TS38.211}
\begin{equation}
Q_u[k]
=
e^{j\alpha_u k}
\bar Q_q[k],
\qquad
k=0,\ldots,M_{\mathrm{SRS}}-1,
\end{equation}

where the base sequence is
\begin{equation}
\bar Q_q[k]
=
e^{-j\frac{\pi qk(k+1)}{N_{ZC}}},
\end{equation}

with $N_{ZC}$ denoting the largest prime satisfying
$N_{ZC}<M_{\mathrm{SRS}}$.
In other words, $\bar{Q}_q[k]$ is the cyclic extension of the corresponding \gls{zc} sequence of length $N_{ZC}$. \gls{zc} sequences of prime length are \gls{cazac}, meaning they exhibit constant amplitude and zero autocorrelation, and their \gls{idft} results in another \gls{zc} sequence with the same properties. However, the cyclic extension partially degrades these ideal characteristics.
Each UE $u$ is assigned a cyclic-shift index
$\eta_u\in\{0,\ldots,7\}$,
which determines the applied phase increment
$\alpha_u
=
\frac{2\pi\eta_u}{8}.$
This phase increment corresponds to a cyclic shift in the time domain of
$N\eta_u/8$
samples, where $N$ denotes the IDFT size.\footnote{A single antenna-port SRS transmission from each UE is considered.}

The subcarrier allocation of $K_{TC}$ ensures that, after the \gls{idft}, the time domain signal consists of $K_{TC}$ identical halves, each lasting half a symbol. Therefore, the time domain waveform $x_{u}[n]$, under the $K_{TC} = 2$ assumption including the \gls{cp}, can be expressed as
\begin{equation}
\label{eq:tx_signal}
x_{u}[n] =
\begin{cases}
s_{u}[n+L], & -L_{CP} \leq n < 0, \\[4pt]
s_{u}[n], & 0 \leq n < L, \\[4pt]
s_{u}[n - L], & L \leq n < N, \\[4pt]
0, & \text{otherwise},
\end{cases}
\end{equation}
where $L_{CP}$ is the \gls{cp} length, $L = N/K_{TC}$, and 
\begin{equation}
    s_{u}[n] = \frac{1}{N}\sum_{k = 0}^{M_{SRS}-1}Q_{u}[k]e^{j\frac{2\pi}{N}(kK_{TC} + k_0)n}.
\end{equation}

The visual representation of \eqref{eq:tx_signal} is shown in the right portion of Fig. \ref{fig:srs_grid}. 
It is important to note that, due to the \gls{cazac} properties of \gls{zc} sequences, two signals with different cyclic shifts applied \( s_{u_i}[n] \) and \( s_{u_j}[n] \) satisfy
\begin{equation}
  \sum_{n=0}^{L-1} s_{u_i}[n]\, s^{\ast}_{u_j}[n] \approx 0, \qquad\text{when} \,\, i \ne j.
\end{equation}
Before the \gls{uav} mission starts, the network partitions with the $C_{SRS}$ and $B_{SRS}$ parameters the available SRS bandwidth into a set of mutually exclusive subbands with the minimum length of 4 \glspl{rb}
$\mathcal{B} = \{ \mathcal{K}_1, \mathcal{K}_2, \dots, \mathcal{K}_{B} \}$,
where each subband \( \mathcal{K}_b = \{ k_{0,b} + m K_{TC} \;|\; m = 0, 1, \dots, M_{\mathrm{SRS}} - 1 \} \)
corresponds to a contiguous set of frequency resources \cite{TS38.211}. These subbands are disjoint, i.e.,
$\mathcal{K}_b \cap \mathcal{K}_{b'} = \emptyset, \quad \forall b \neq b'$,
and their union spans the allocated SRS bandwidth $\mathcal{B}$.

Each subband \( \mathcal{K}_b \) can be assigned to one or more \glspl{ue}, denoted by the set
\( \mathcal{U}_b \subseteq \mathcal{U} \),
where every \gls{ue} is characterized by a unique cyclic shift index \( \eta_u \) within that subband.
The SRS periodicity \( T_{\mathrm{per}} \) is set to be the same for all \glspl{ue},
and all SRS-related configuration parameters are communicated to the \gls{uav} at the beginning of the mission. We remark that each \gls{ue} is associated with a unique pair \( (\mathcal{K}_b, \eta_u) \), which uniquely identifies the \gls{ue}.

In practical deployments, we assume that the SRS configuration (e.g., subband allocation and cyclic shifts) is predefined prior to the mission within a tactical wireless bubble, where the network is fully controlled and can enforce specific \gls{ul} configurations. These parameters are assigned to the UEs using standard 3GPP 5G configuration procedures and are made available to the sensing UAV before the mission starts \cite{TS38.211, 3gpp38331_rel18_v18_8_0, 3gpp38213_rel18_v18_8_0}. During operation, no further coordination with the network is required, and the UAV passively exploits the resulting \gls{ul} transmissions.

\subsection{Received Signal Model}

The signal received at the  \gls{uav}, located at position \( \mathbf{p}[r] \) at the $r$th time step, on its antenna \( \ell \), from the set of \glspl{ue} assigned to subband \( \mathcal{K}_b \), \( u \in \mathcal{U}_b = \{ I_j \in \mathcal{U} : I_j \rightarrow \mathcal{K}_b \} \), each experiencing a set of propagation paths \( \mathcal{P}_{u,\ell} = \{1, \ldots, P_{u,\ell}\} \), can be expressed in the time domain as \cite{goldsmith2005wireless, TS38.211}
\begin{equation}
y_{\ell}(t)\hspace{-2pt}=\hspace{-3pt}\sum_{u=1}^{U_b}\hspace{-2pt}y_{u,\ell}(t)\hspace{-2pt}=\hspace{-3pt}\sum_{u=1}^{U_b}\hspace{-2pt}\sum_{p=1}^{P_{u,\ell}}\!
\hspace{-3pt}a_{p,u,\ell}z_{u}(t\!-\!\tau_{p,u,\ell})e^{j2\pi(\nu_{p,u,\ell}\!+\!\nu_{u,\mathrm{cfo}})t},
\label{eq:Rx_signal}
\end{equation}
where \( U_b = |\mathcal{U}_b| \) denotes the number of transmitting \glspl{ue} in subband \( \mathcal{K}_b \), $P_{u,\ell}$ the number of propagation paths of \gls{ue} $u$, and $z_u(t)$ its transmitted signal. The \gls{toa} of the $p$th path is $\tau_{p,u,\ell} $$\,=\,$$ \tau'_{p,u,\ell} $$\,-\,$$ TA_u$, where $\tau'_{p,u,\ell}$ is the physical propagation delay and $TA_u$ is the applied timing advance. For each \gls{ue}, the propagation delays are ordered in ascending order, i.e.,
$\tau_{1,u} < \tau_{2,u} < \cdots < \tau_{P_{u,\ell},u}.$
 The Doppler shift and \gls{cfo} are denoted by $\nu_{p,u,\ell}$ and $\nu_{u,\mathrm{cfo}}$, respectively, while the complex path gain is $a_{p,u, \ell} $$\,=\,$$ \tilde a_{p,u,\ell} e^{j2\pi f_c \tau'_{p,u,\ell}}$, with $\tilde a_{p,u,\ell}$ the complex baseband channel coefficient and $f_c$ the carrier frequency. All these parameters, except \gls{cfo}, implicitly depend on the \gls{uav} position $\mathbf{p}[r]$. For simplicity in the notation, we will, from now on, consider the received signal at a single antenna and a fixed \gls{srs} subband, omitting the subscripts $\ell$ and $b$.

\textbf{Remark}. In the following, we consider multiple UEs multiplexed within the smallest SRS subband $\mathcal{K}_b$ (1.4\,MHz) using cyclic shifts, as this minimizes sampling rate and onboard processing requirements, which are critical for lightweight UAV platforms. This corresponds to a worst-case condition in terms of intra-band interference. More generally, the SRS structure enables high multiplexing capability by combining cyclic shifts within each subband and frequency multiplexing across multiple subbands, allowing interference-free operation at the cost of increased bandwidth and processing complexity.

\section{UEs synchronization, identification, and localization} 
\label{sec:workflow}

In this section, we describe the signal processing procedures that enable the sensing \gls{uav} to autonomously synchronize, identify, and localize multiple \glspl{ue} from their \gls{ul} \glspl{srs}. 
The synchronization and \gls{ue} identification stages run independently on each antenna of the \gls{uav}, while localization exploits antenna diversity by combining outputs from all antennas.

\subsection{Coarse Time Synchronization via SRS Structure} \label{sec:Coarse time sync}

To obtain a coarse time alignment, we exploit the inherent repetition structure of the \gls{srs}. In particular, by adopting the SRS configuration described in Sec. \ref{subsec:SRS} with $K_{TC}=2$, the transmitted signal exhibits a structured time-domain pattern that is well suited for synchronization.
 
We consider the detection metric \cite{650240}
\begin{equation}
M[n] = \frac{|P[n]|^2}{R[n]^2},
\end{equation}
where
\begin{equation}
\begin{aligned}
P[n] &= \sum_{l = 0}^{L-1} y[n + l]\, y^*[n+l+L],\\
R[n] &= \frac{1}{2} \Biggl( \sum_{l = 0}^{L-1} |y[n + l]|^2 + \sum_{l = 0}^{L-1} |y[n + l + L]|^2 \Biggr).
\end{aligned}
\end{equation}
The metric \(M[n]\) is bounded between 0 and 1. 
This procedure provides coarse synchronization with any detected \gls{srs}, leveraging their unique repeated time-domain structure imposed by setting $K_{TC} = 2$ in the protocol configuration. It is robust to \gls{cfo}, as phase rotations cancel out in $P[n]$, and can be efficiently computed using recursive updates.

\begin{figure}[!t]
    \centering
    \subfloat[Behavior of $f_0$ and $g_0$ as functions of the relative \protect\gls{srs} arrival time $\Delta\tau_{p,u}$.]{
        \includegraphics[width=0.43\linewidth]{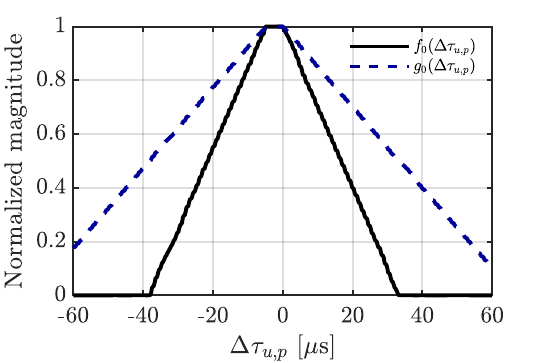}
        \label{fig:f0_g0}
    }
    \hfill
    \subfloat[Detection metric when the condition in \protect\eqref{eq: high metric condition} holds $\forall p,u$ (blue), and when it does not (black).]{
        \includegraphics[width=0.43\linewidth]{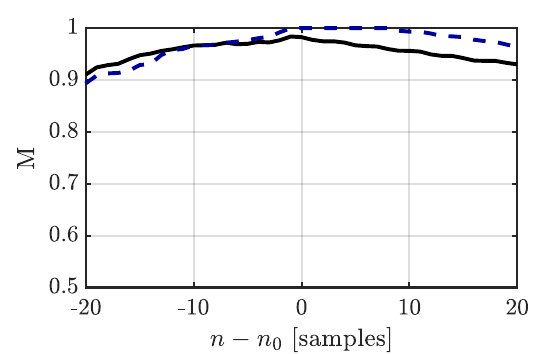}
        \label{fig:M_with_without_CP}
    }
    \caption{Impact of delay spread on the synchronization detection metric.}
    \label{fig:metric_and_functions}
\end{figure}

The \glspl{toa} of the \glspl{srs} received at the \gls{uav} from different \glspl{ue} are generally mutually misaligned, since the \gls{uav} is not located at the \gls{gnb} site where the \gls{ta} aligns the streams. 
For convenience, we select one arrival time $\tau_{\bar p,\bar u}$ as reference, with $\bar u \in \mathcal{U}$ and $\bar p \in \mathcal{P}_{\bar u}$, and express all others relative to it
\begin{equation}
    \Delta\tau_{p,u} = \tau_{p,u} - \tau_{\bar p,\bar u}.
\end{equation}
To reliably detect the presence of \glspl{srs}, the metric should have values close to unity (i.e., $M[n]\approx 1$) at sufficiently high \gls{snr}. 
Evaluating \(M[n]\) at the reference arrival index
\(n_0=\lceil\tau_{\bar p,\bar u}/T_s\rceil\), and exploiting the quasi-orthogonality of different SRS signals together with the piecewise approximation of the correlation and energy terms (see Appendix~\ref{app: expansion of detection metric} for the complete derivation), the synchronization metric can be approximated as
\begin{equation}
\label{eq: M_n0}
M[n_0] \approx
\left(
\frac{\displaystyle\sum_{u=1}^{U}\sum_{p=1}^{P_u}
|a_{p,u}|^2\,f_0(\Delta\tau_{p,u})}
{\displaystyle\sum_{u=1}^{U}\sum_{p=1}^{P_u}
|a_{p,u}|^2\,g_0(\Delta\tau_{p,u})}
\right)^2.
\end{equation}
where $f_0(\Delta\tau_{p,u}), g_0(\Delta\tau_{p,u}) \in [0,1]$ depend on the temporal misalignment $\Delta\tau_{p,u}$. 
Figure~\ref{fig:f0_g0} shows that both satisfy $f_0 = g_0 = 1$ whenever
\begin{equation} \label{eq: high metric condition}
    -T_{\mathrm{CP}} \le \Delta\tau_{p,u} \le 0,
\end{equation}
with $T_{\mathrm{CP}}$ denoting the cyclic prefix duration. 
Outside this interval, it holds that $g_0(\Delta\tau_{p,u}) $$\,>\,$$ f_0(\Delta\tau_{p,u})$, reducing $M[n_0]$. Figure~\ref{fig:M_with_without_CP} illustrates the two outcomes. 
If the relative arrival times $\Delta\tau_{p,u} \ \forall\, p,u$ lie within the interval defined in \eqref{eq: high metric condition}, the metric attains its maximum value (blue curve) and remains 1 until at least one arrival time falls outside this interval.
Conversely, if the condition is never satisfied, the metric remains strictly below 1 (black curve).

In practice, signals from distant \glspl{ue} typically exhibit much lower power than those from nearby \glspl{ue}, i.e., 
$|a_{p,u_{\mathrm{near}}}|^2 \gg |a_{p,u_{\mathrm{far}}}|^2$. 
Since distant \glspl{ue} also induce larger propagation delays, they are the primary contributors to potential degradation of the synchronization metric in \eqref{eq: M_n0}. 
However, their impact is often mitigated by significant path-loss attenuation, which reduces their received power contribution. 
In 5G, the SRS transmit power is governed by standard \gls{ul} power-control procedures configured via \gls{rrc} signaling \cite{3gpp38213_rel18_v18_8_0, 3gpp38331_rel18_v18_8_0}. 
By assigning identical SRS power-control parameters to all UEs, the network can enforce a uniform transmit power policy, so that variations in received power at the UAV predominantly reflect large-scale propagation effects. 
Under this configuration, the received power naturally correlates with the UE--UAV distance, ensuring that distant \glspl{ue} are inherently received with substantially lower power than nearby ones, thereby limiting their contribution to the synchronization metric and improving robustness to multi-\gls{ue} interference.

Once the synchronization metric \( M[n] \) is computed, the sensing \gls{uav} must identify the point at which it reaches its maximum value and exceeds a predefined threshold \( M_{\mathrm{th}} \). 
This point provides a coarse estimate of the \gls{toa} of the received \glspl{srs} and allows the receiver to achieve an initial time alignment. When received \glspl{srs} are synchronized, the timing metric exhibits a sharp rise followed by a plateau whose duration equals the \gls{cp} length $L_{CP}$. This behavior arises because, when $n$ is within the cyclic prefix interval, the two identical halves of the preamble align perfectly, contributing coherently to the correlation and yielding a constant maximum value of the metric rather than a single narrow peak. Therefore, to improve the detection, the metric \( M[n] \) is convolved with a rectangular window of length equal to $L_{CP}$.

The initial detection instant $n_{sync}$ is determined by 
\begin{equation}
\hspace{-0.1cm}M[n_{\mathrm{sync}}] \hspace{-0.1cm}\ge \hspace{-0.1cm}M_{\mathrm{th}}
\hspace{-0.1cm}\;\land\;
\hspace{-0.1cm}D[n_{\mathrm{sync}} \hspace{-0.1cm}- 1]\hspace{-0.1cm} >\hspace{-0.1cm} \delta
\hspace{-0.1cm}\;\land\;\hspace{-0.1cm}
D[n_{\mathrm{sync}}] \hspace{-0.1cm}< \hspace{-0.1cm}-\delta,
\end{equation}
where $\delta$ is a small positive quantity and  $D[n] = M[n+1] - M[n]$ is the discrete derivative of the metric. For moderate \glspl{snr} (\( < 20 \)~dB) within the investigated subband, and when \eqref{eq: high metric condition} holds $\forall p,u$, the \gls{snr} can be estimated as
\begin{equation}
\label{eq: SNR}
    \hat{\mathrm{SNR}} \approx \frac{\sqrt{M(n_{sync})}}{1 - \sqrt{M(n_{sync})}}.
\end{equation}
Consequently, the detection threshold \( M_{\mathrm{th}} \) can be set based on the minimum \gls{snr} level that must be reliably detected \cite{650240}.

\subsection{Fine Synchronization and UE Identification}

Once this initial timing is established, the \gls{uav} must: (i) separate the incoming signals, (ii) identify the transmitting \glspl{ue}, (iii) localize them, (iv) refine the synchronization error.  
Specifically, the residual synchronization error for \gls{ue} $u$ and path $p$ is defined as
\begin{equation}
    \varepsilon_{p,u} = n_{p,u} - n_{sync},
\end{equation}
where $n_{sync}$ denotes the initial coarse synchronization instant and $n_{p,u} = \lceil \tau_{p,u}/T_s \rfloor $ the actual \gls{toa}.

\subsubsection{Separation of \gls{ul} Signals} This step consists of discriminating the individual \gls{ue} signals within the received composite waveform, thereby enabling \gls{ue} identification and subsequent processing. 

The received \gls{srs} signal can be expressed as
\begin{equation}
\begin{aligned}
     y_{\text{SRS}}[n]
    = \sum_{u=1}^{U} \sum_{p=1}^{P_u} 
     \hspace{-0.15cm}a_{p,u}
    x_u\!\left[n - L_{\text{CP}} - \varepsilon_{p,u} \right] 
    e^{j 2\pi (\nu_{p,u} + \nu_{u,\text{cfo}}) \, n T_s},
\end{aligned}
\end{equation}
with $n = 0, \dots, L-1$.
Exploiting the repeated time structure, this can be rewritten (neglecting a common phase term) as
\begin{equation}
\begin{aligned}
    y_{\text{SRS}}[n] 
    \hspace{-0.05cm}= \hspace{-0.05cm}\sum_{u=1}^{U} \sum_{p=1}^{P_u} 
    \hspace{-0.05cm}a_{p,u} \,
    s_u\!\bigl[(n - \varepsilon_{p,u}) \bmod L \bigr]
    e^{j 2\pi (\nu_{p,u} + \nu_{u,\text{cfo}}) \, n T_s}.
\end{aligned}
\end{equation}
Applying an $L$-point \gls{dft}, we obtain the received spectrum 
$Y_{\text{SRS}}[k']$ for $k' $$\,=\,$$0,  \dots, L$$\,-\,$$1$. We index the $M_{\text{SRS}}$ active subcarriers through the mapping
$\kappa[m] \triangleq \frac{k_{0}+mK_{\mathrm{TC}}}{K_{\mathrm{TC}}}, \qquad m=0,\dots,M_{\mathrm{SRS}}-1,$
such that the received spectrum on the $m$th active subcarrier is 
\begin{equation}
    \hspace{-0.15cm}Y_{\text{SRS}}[m] 
   \hspace{-0.1cm} = \hspace{-0.1cm}Y_{\text{SRS}}(\kappa[m])
    \hspace{-0.1cm}= \hspace{-0.15cm}\sum_{u=1}^{U}\hspace{-0.1cm}\sum_{p=1}^{P_u} 
    \hspace{-0.15cm}a_{p,u} S\big(m\big) 
    e^{j 2\pi m \left( \frac{\eta_u}{8} - \frac{\varepsilon_{p,u}}{L} \right)}.
\end{equation}
We then multiply the received spectrum by the known reference sequence
\begin{equation}
    c[m] = Y_{\text{SRS}}[m] \, S^*[m]=\sum_{u=1}^{U} \sum_{p=1}^{P_u} 
     a_{p,u} \, e^{j 2\pi m f_{p,u}},
\end{equation}
where  $ f_{p,u} = {\eta_u}/{8} - {\varepsilon_{p,u}}/{L}$.
This compact expression shows that each \gls{ue}--path pair received by the \gls{uav} manifests as a complex exponential in the frequency domain, enabling the application of frequency estimation techniques for refined \gls{toa} and cyclic shift estimation. It is worth recalling that each pair \((\mathcal{K}_b, \eta_u)\), representing the allocated SRS bandwidth and cyclic shift respectively, uniquely identifies a single \gls{ue}. Therefore, by estimating the received cyclic shift and determining the corresponding occupied bandwidth, the \gls{uav} can discriminate among the active \glspl{ue} and associate each received signal with its transmitting \gls{ue}.

\begin{figure}[!t]
    \centering
    \includegraphics[width=0.9\linewidth]{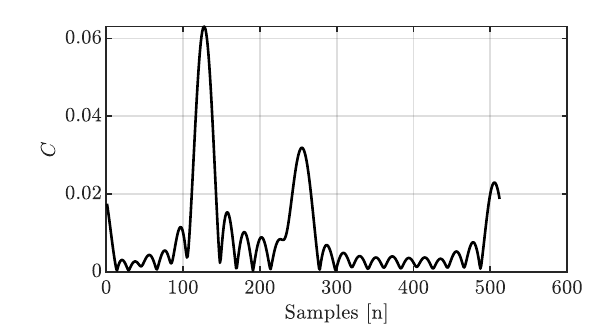}
    \caption{Example of the computed \( C[k] \) for three different \glspl{ue} applying cyclic shifts 0, 4, and 2, respectively.}
    \label{fig: C[k]}
\end{figure}

Applying an $L$-point \gls{dft} on $c[m]$ yields
\begin{equation}
\hspace{-0.1cm}C[k]
\hspace{-0.1cm}= \hspace{-0.1cm}\sum_{u,p}
\hspace{-0.1cm}a_{p,u}\,
\hspace{-0.1cm}e^{-(j\,2\pi\,\chi_{p,u}[k]\,(M_{\mathrm{SRS}}-1))}
\hspace{-0.05cm}\frac{\sin\!\big(\pi M_{\mathrm{SRS}}\chi_{p,u}[k]\big)}{\sin\!\big(\pi \chi_{p,u}[k]\big)},
\end{equation}
where $\chi_{p,u}[k] = f_{p,u} - {k}/{L}$.
Figure~\ref{fig: C[k]} illustrates an example of the computed 
$C[k]$ obtained from \gls{srs} signals received by three different \glspl{ue} applying different cyclic shifts.

\begin{algorithm}[!b] 
\fontsize{8}{8}\selectfont
\caption{Matching Pursuit over \gls{dft} dictionary}
\label{alg:matching_pursuit}
\begin{algorithmic}[1]
\Require SRS channel $c[m]$, half symbol length $L$, $M_{\mathrm{SRS}}$
\State Initialize $c^{(0)}[m] = c[m]$, $i=1$
\While{BIC decreases}
    \State Compute \gls{dft}: $C^{(i)}[k] = FFT_L(c^{(i-1)}[m])$
    \State Find peak: $\hat{k}^{(i)} = \arg\max_k |C^{(i)}[k]|$
    \State Estimate: $\hat{f}^{(i)} = \hat{k}^{(i)}/L$
    \State Estimate: $\hat{a}^{(i)} = \frac{1}{M_{\mathrm{SRS}}}
    \sum_{m=0}^{M_{\mathrm{SRS}}-1} c^{(i-1)}[m] e^{-j 2\pi m \hat{f}^{(i)}}$
     \State Estimate : $\hat{\gamma}^{(i)} = 20 \log_{10} |\hat{a}^{(i)}|$
    \State Update: $c^{(i)}[m] = c^{(i-1)}[m] - \hat{a}^{(i)} e^{j 2\pi m \hat{f}^{(i)}}$
    \State Compute BIC$^{(i)}$ as in \eqref{eq:BIC}
    \State $i \gets i + 1$
\EndWhile
\end{algorithmic}
\end{algorithm}

\subsubsection{Peaks Extraction}
\label{sec: peaks extraction}
The dominant frequency components of the received signal are iteratively extracted using a Matching Pursuit (MP) algorithm \cite{wang2012generalized} over a sinusoidal dictionary, i.e. the \gls{dft} bins. 
If the dictionary resolution is coarse, oversampling and parabolic interpolation refine the estimates. 
At each iteration, the strongest component is removed from the residual, and its amplitude and equivalent power are estimated. 
The process continues until the \gls{bic} indicates that adding further components no longer improves the model.
The \gls{bic} is computed as  \cite{neath2012bayesian}
\begin{equation} \label{eq:BIC}
\hspace{-0.17cm}\text{BIC}^{(i)} \hspace{-0.1cm}=\, \hspace{-0.1cm}2i \log M_{\mathrm{SRS}} 
\hspace{-0.1cm}+ \hspace{-0.1cm}M_{\mathrm{SRS}} \log \!\left( 
\hspace{-0.1cm}\frac{1}{M_{\mathrm{SRS}}} \hspace{-0.2cm}\sum_{m=0}^{M_{\mathrm{SRS}}-1} \hspace{-0.2cm}|c^{(i)}[m]|^2 
\hspace{-0.1cm}\right).
\end{equation}
The procedure is summarized in Algorithm~\ref{alg:matching_pursuit}.

\subsubsection{User Identification via Cyclic Shift}
\label{sec: user identification}

Note that, due to the different cyclic shifts applied, the frequencies associated with different \glspl{ue} are well separated, which facilitates their extraction. 
We can convert the estimated frequencies $\hat{f}^{(i)}$ into the corresponding estimated cyclic shift indices as
\begin{equation}
\hat{\eta}^{(i)}
=
\left\lfloor
U\,
\operatorname{mod}\!\left(\hat f^{(i)} + \frac{1}{2\,U},\, 1\right),
\right\rfloor
\end{equation}
where $U$ denotes the total number of cyclic shifts assigned to the \glspl{ue} (i.e. the number of \glspl{ue}) within the considered band.
If the cyclic shifts are uniformly (i.e., maximally) spaced and $U$ is even, to avoid a misdetection of the \gls{ue}–path tuple $(u,p)$, the synchronization error $\varepsilon_{p,u}$, must satisfy
\begin{equation}
\label{eq: error_bounds}
    \varepsilon_{p,u} \in \left( -L/(2 U), \, L/(2U) \right] .
\end{equation}
Consequently, our goal is to maximize $L$$\,=\,$$N/K_{\mathrm{TC}}$ by minimizing $K_{\mathrm{TC}}$. Therefore, the optimal SRS subcarrier step is $K_{\mathrm{TC}} $$\,=\,$$ 2$, its minimum. Furthermore, depending on the scenario size and UAV processing capabilities, fewer cyclic shifts than the maximum per band (i.e., $ U_{\mathrm{max}} $$\,=\,$$ 8 $) can be allocated, reducing \gls{ue} identification errors.
Once the cyclic shift index \(\hat{\eta}^{(i)}\) is estimated, the identity of the transmitting \gls{ue} can be unambiguously determined by pairing it with the corresponding frequency band \(\mathcal{K}_b\).

\subsubsection{Fine synchronization}
The timing synchronization described in Sec.~\ref{sec:Coarse time sync} allows us to coarsely synchronize to the received \gls{srs} signals. 
This operation is required only once, since the \gls{srs} periodicity of the transmitting \glspl{ue} is known by the \gls{uav}. 
This allows us to maintain synchronization at subsequent receptions without recomputing the timing metric from scratch every \gls{srs} reception. 

Starting from the initial synchronization instant \( n_{sync} \), the synchronization points are updated at each \gls{srs} reception as
\begin{equation}
    n_q = n_{sync} + q T_{\mathrm{per}} - \tilde{\varepsilon}(q),
\end{equation}
where \( q \in \mathbb{N} \), \( T_{\mathrm{per}} \) is the \gls{srs} periodicity and \( \tilde{\varepsilon}(q) \) is the filtered residual synchronization error, updated through
\begin{equation}
    \tilde{\varepsilon}(q) = 
    \beta \, \hat{\varepsilon}_{\bar p, \bar u}(q)
    + (1 - \beta)\, \tilde{\varepsilon}(q-1),
\end{equation}
with \( \beta \in [0,1] \) controlling the weighting between the latest and previous estimates.

The residual synchronization error arises for two main reasons. 
First, due to noise and mutual \gls{srs} interference, the initial \( n_{sync} \) is typically not perfectly aligned with the \gls{srs} peak but falls somewhere between replicas. 
Second, even after fine synchronization, the \gls{uav}'s and \glspl{ue}' clock drift over time, introducing an increasing offset that must be compensated.

Since fine synchronization can only be accurately aligned with respect to a single received \gls{srs} (the others will remain slightly misaligned, except when the \gls{uav} is located in correspondence with the \gls{gnb} site), the most powerful received \gls{srs} is typically chosen as the reference. 
The residual error associated with this reference path at the $q$th \gls{srs} reception is then estimated as
\begin{equation}
    \hat \varepsilon_{\bar p, \bar u}(q) 
    = \left\lfloor L \bigl( \hat f^{(1)}(q) - \hat \eta^{(1)}(q)/8 \bigr) \right\rfloor,
\end{equation}
where the first iteration of the frequency estimation process always corresponds to the strongest received replica.

\subsection{Measurement Cleaning and Detection Reliability}
\label{sec:measurement_cleaning_fp}

After frequency estimation, the UAV obtains vectors of estimated metrics
\(\tilde{\boldsymbol{\gamma}} = [ \hat \gamma^{(1)}, \dots, \hat \gamma^{(i_{end})} ]^T\)
and corresponding cyclic shifts indexes
\(\tilde{\boldsymbol{\eta}} = [ \hat \eta^{(1)}, \dots, \hat \eta^{(i_{end})} ]^T\),
where each element represents a detected \gls{ue}-path pair \((u,p)\), and $\hat \gamma^{(j)}$ denotes the strength of the corresponding SRS component. The index $i_{end}$ is the maximum number of iterations.

To prepare for localization, only the strongest component per UE is retained. This choice is consistent with the considered narrowband setting, where resolving individual multipath components is not reliable, and the dominant component provides the most stable and informative contribution for localization.
For each \gls{ue} \( u \in \mathcal{U}_b \), define
$\mathcal{J}_u = \{ j \mid \hat{\eta}^{(j)} = \eta_u \}, \hat \gamma_u^{\max} = \max_{j \in \mathcal{J}_u} \hat \gamma^{(j)}$,
and discard all other replicas. The cleaned measurement vector is
$\boldsymbol{\gamma}_{\mathrm{SRS}} = [\, \hat \gamma_1^{\max}, \dots, \hat \gamma_{U}^{\max} \,]$.
Missed detections due to low received power can be flagged for subsequent processing.

The behavior of the proposed detection algorithm under the main failure mechanisms is summarized below.
At low \gls{snr}, the synchronization metric \(M[n]\) rarely exceeds the detection threshold \(M_{\mathrm{th}}\). Consequently, the peak extraction and the subsequent MP stages are typically not activated, resulting in a missed detection.
False positives may instead occur when unrelated narrowband signals exceed \(M_{\mathrm{th}}\). This typically happens when their envelope varies slowly over the observation interval, preserving a significant correlation between the two halves of the synchronization window and producing large values of the synchronization metric. Whenever \(M_{\mathrm{th}}\) is exceeded, the received signal undergoes the nominal SRS preprocessing followed by a second \gls{dft}. Under the SRS hypothesis, the resulting spectrum exhibits a sparse structure with a few well-defined peaks at specific frequency bins (see Fig.~\ref{fig: C[k]}), whereas unrelated signals produce a noise-like spectrum. The \gls{bic}-based stopping criterion exploits this difference: if the extracted components are not supported by the data, the detection is classified as a false positive, discarded, and the synchronization metric is evaluated again on newly acquired samples.
Strong multipath may also reduce the received SRS power because of destructive interference, occasionally leading to missed detections. However, since SRS transmissions are periodic and the UAV continuously acquires measurements while moving, different channel realizations are observed over time, making subsequent SRS transmissions likely to provide valid measurements. Moreover, even if an occasional false positive passes all detection stages, it typically appears as an isolated observation and has a negligible impact on the subsequent UE identification and localization procedures, which rely on multiple SRS measurements collected at different UAV positions.

\begin{algorithm}[!t]
\fontsize{8}{8}\selectfont
\caption{Weighted Mean-Shift Localization for \gls{ue} $u$}
\label{alg:localization}
\begin{algorithmic}[1]
\Require 
 Per-antenna sequences $\{\gamma_{\mathrm{SRS, ant}}^{(u,\ell)}[r]\}$, UAV positions $\{\mathbf{p}_{u,r}\}$, 
threshold $\gamma_{\mathrm{th}}$, kernel bandwidths $h_x,h_y$, tolerance $\delta$, max iterations $i_{\max}$
\Ensure Estimated position $\hat{\mathbf{m}}_u$

\State Select best antenna: 
$\ell_u^\star[r] \gets \arg\max_{\ell \in \mathcal{A}} \gamma_{\mathrm{SRS}}^{(u,\ell)}[r]$
\State Form measurement sequence: 
$\gamma_{\mathrm{SRS}}^{(u)}[r] \gets \gamma_{\mathrm{SRS}}^{(u,\ell_u^\star[r])}[r]$
\State Normalize: $\tilde{\gamma}^{(u)}[r] \gets \gamma_{\mathrm{SRS}}^{(u)}[r] / \max_r \gamma_{\mathrm{SRS}}^{(u)}[r]$
\State Select valid samples: $\mathcal{R}_u \gets \{ r \mid \tilde{\gamma}^{(u)}[r] \ge \gamma_{\mathrm{th}} \}$
\State Set weights: $w_{u,r} \gets \tilde{\gamma}^{(u)}[r]$
\State Initialize $\hat{\mathbf{m}}_u^{(0)}$, $i \gets 0$

\Repeat
    \State Update position $\hat{\mathbf{m}}_u^{(i+1)} \text{as in} ~\eqref{eq:mean_shift_update_final}$
    \State $i \gets i + 1$
\Until{$\|\hat{\mathbf{m}}_u^{(i+1)} - \hat{\mathbf{m}}_u^{(i)}\| < \delta$ \textbf{or} $i \ge i_{\max}$}

\State \Return $\hat{\mathbf{m}}_u \gets \hat{\mathbf{m}}_u^{(i+1)}$
\end{algorithmic}
\end{algorithm}
\subsection{UEs Localization}
\label{sec:localization}

The accurate estimation of \gls{ue} positions is fundamentally based on the SRS-specific metric \(\gamma_{SRS}[r]\) collected for each measurement point during the identification process.  
Since each UAV is equipped with \(N_{ant}\) antennas, at every measurement instant \(r\) the receiver obtains, for each \gls{ue} \(u\), the antenna-specific metric vector
$\boldsymbol{\gamma}_{\mathrm{SRS},ant}^{(u)}[r] =
\big[\,\gamma_{\mathrm{SRS}}^{(u,1)}[r],~\dots,~\gamma_{\mathrm{SRS}}^{(u,N_{ant})}[r]\,\big]^T$.

To reduce antenna-dependent polarization and small-scale fading fluctuations while retaining the large-scale spatial trend relevant for localization, we select, for each \gls{ue} $u$, at the measurement instant $r$, the antenna whose \gls{srs}-derived metric exhibits the highest metric
\begin{equation}
\label{eq:antenna_selection}
\ell_u^\star[r] = \arg\max_{\ell \in \mathcal{A}} ~ \gamma_{\mathrm{SRS},ant}^{(u,\ell)}[r],
\end{equation}
and define the selected measurement sequence as
$\gamma_{\mathrm{SRS}}^{(u)}[r] \triangleq \gamma_{\mathrm{SRS},ant}^{(u,\ell_u^\star[r])}[r]$.
By stacking the selected sequences over time, we obtain
$\boldsymbol{\Gamma}_{\mathrm{SRS}} =
\begin{bmatrix}
\boldsymbol{\gamma}_{\mathrm{SRS}}^{(1)} &
\cdots &
\boldsymbol{\gamma}_{\mathrm{SRS}}^{(U)}
\end{bmatrix}^{\!T}
\in \mathbb{R}^{U \times R}.$

Each row is normalized independently:
\begin{equation}
\tilde{\boldsymbol{\gamma}}_{\mathrm{SRS}}^{(u)}
=
\frac{\boldsymbol{\gamma}_{\mathrm{SRS}}^{(u)}}{\max_{r}\gamma_{\mathrm{SRS}}^{(u)}[r]},
\qquad u = 1,\dots,U,
\end{equation}
and unreliable measurements are discarded using
$\mathcal{R}_u
=
\{\, r \mid \tilde{\gamma}_{\mathrm{SRS}}^{(u)}[r] \ge \gamma_{\mathrm{th}} \,\}.$ Each valid sample is then mapped to the UAV position and localization weight:
$\mathbf{p}_{u,r} = \mathbf{p}^{(2D)}[r], \quad w_{u,r} = \tilde{\gamma}_{\mathrm{SRS}}^{(u)}[r]$.

The localization is performed using a weighted mean-shift procedure \cite{wu2007mean}
\begin{equation} \label{eq:mean_shift_update_final}
\hspace{-0.2cm}\hat{\mathbf{m}}_u^{(i+1)} =
\frac{
\displaystyle
\sum_{r \in \mathcal{R}_u} 
\hspace{-0.2cm}\mathbf{p}_{u,r} w_{u,r} 
e^{-\frac{1}{2} 
(\mathbf{p}_{u,r}-\hat{\mathbf{m}}_u^{(i)})^\top 
\mathbf{H}^{-1} 
(\mathbf{p}_{u,r}-\hat{\mathbf{m}}_u^{(i)})}
}{
\displaystyle
\sum_{r \in \mathcal{R}_u} 
\hspace{-0.2cm}w_{u,r} 
e^{-\frac{1}{2} 
(\mathbf{p}_{u,r}-\hat{\mathbf{m}}_u^{(i)})^\top 
\mathbf{H}^{-1} 
(\mathbf{p}_{u,r}-\hat{\mathbf{m}}_u^{(i)})}
}
\end{equation}
where
$\mathbf{H} = \operatorname{diag}\!\left([h_x^2,\; h_y^2]^T\right)$, and $h_x$ and $h_y$ denote the kernel bandwidths along the East--West and North--South directions, respectively. The algorithm iterates until 
$\|\hat{\mathbf{m}}_u^{(i+1)} - \hat{\mathbf{m}}_u^{(i)}\| < \delta,$
and the final estimate is denoted by \(\hat{\mathbf{m}}_u\). A Gaussian kernel is adopted due to its smoothness and infinite support, which ensure a continuously differentiable spatial density estimate and stable fixed-point updates in the mean-shift iteration.

During the localization process, the \gls{uav} first moves to the center of the designated polygonal search area and then executes an initial perimeter scan.  
The collected measurements are processed onboard in real time to provide a first coarse estimate for each \gls{ue} position.  
The UAV then transitions to a refinement phase, during which it performs a small-scale, high-resolution search (typically hexagonal) centered around each preliminary estimate, with navigation waypoints continuously updated based on the estimated \gls{ue} positions.  
This phase yields dense measurement clusters that further improve localization accuracy.  
Once all \glspl{ue} are refined, the UAV transmits the final estimates to the \gls{gcs} and executes a return-to-home sequence, completing the multi-\gls{ue} localization mission. The framework assumes quasi-static UEs during each localization cycle, i.e., the UE position is considered constant over the duration of the UAV maneuver. 
This assumption is consistent with emergency scenarios, where individuals in distress are typically stationary for a few minutes.

The proposed localization workflow is summarized in Alg.~\ref{alg:localization}.
\begin{figure}[!t]
    \centering
    \includegraphics[width=\linewidth]{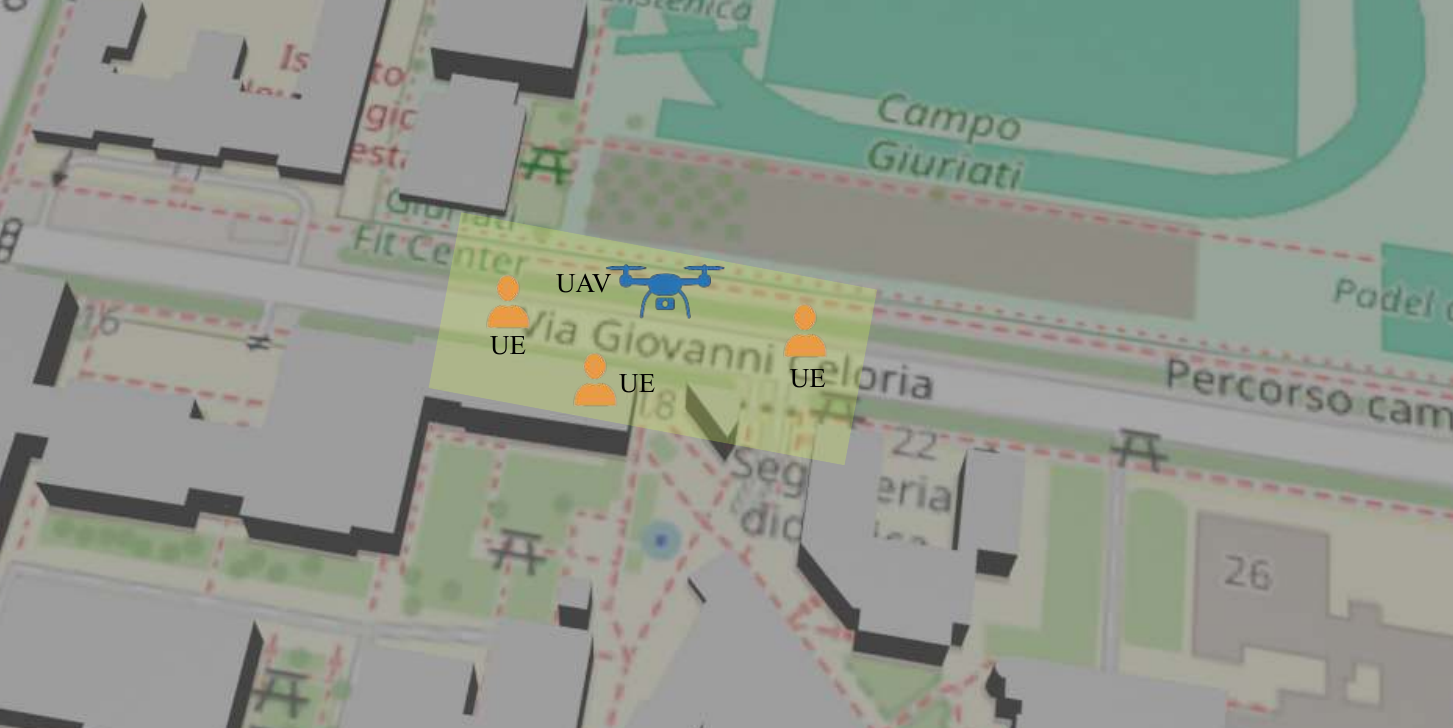}
    \caption{3D view of the considered urban scenario during simulations.  In green, the mission area.}
    \label{fig:urban_scenario_simulation}
\end{figure}

\section{Simulation Results} \label{sec:results}

We report here the numerical results obtained from the evaluation of the proposed framework for both \glspl{ue} synchronization, identification and localization. 

\subsection{Scenario and LAWN settings}
A realistic environment was reconstructed from \textit{OpenStreetMap}. 
We considered two scenarios: (i) an urban area near the Giuriati Sports Campus in Milan and (ii) a rural environment, both covering approximately $120\,\text{m} \times 80\,\text{m}$ (see Fig.~\ref{fig:urban_scenario_simulation}), which is compatible with the operational time constraints of lightweight UAV platforms. The geographic data were imported into \textsc{Matlab} to extract scenario boundaries and building heights.
LoS/NLoS conditions between the UAV and the UEs were computed using the \textsc{Matlab} ray-tracing toolbox, and channel realizations were generated with the \textsc{QuaDRiGa} library. The \texttt{3GPP\_38.901\_RMa\_LOS/NLOS} and \texttt{3GPP\_38.901\_UMa\_LOS/NLOS} models were adopted for rural and urban scenarios, respectively, providing a reasonable approximation for UAV communications up to 25\,m altitude. Doppler effects were included based on UAV motion, with a maximum speed of $6\,\text{m/s}$. UE antennas were modeled as randomly oriented half-wave dipoles.

The UAV was equipped with two ADALM-PLUTO SDR antennas \cite{plutoSDR_wiki}, spaced $30\,\text{cm}$ apart and oriented orthogonally. Their 3D radiation pattern was approximated from measured horizontal and vertical cuts. The receiver noise figure was set to $7\,\text{dB}$ and the carrier frequency to $2.4\,\text{GHz}$. The detection and synchronization thresholds were fixed to $\gamma_{th} = M_{th} = 0.6$. The kernel bandwidths $h_x$ and $h_y$ were set to half of the width and height of the spatial region of interest, respectively, in order to operate at a global spatial scale and avoid spurious local maxima due to noisy measurements.
The UAV flight grid was discretized with a spatial resolution of $1\,\text{m}$ and a fixed altitude of $25\,\text{m}$, chosen as a trade-off between localization sensitivity and operational feasibility, as higher altitudes reduce the horizontal power gradient while lower ones are often constrained by obstacles. Uplink traffic was emulated using SDR recordings with active narrowband SRS generated via the \gls{oai} platform, configured according to the proposed optimized parameters for synchronization and identification. To model realistic hardware effects, a UE clock drift of $0.05$\,ppm was introduced with periodic compensation \cite{ JSAC_Paglierani}.

\subsection{UEs Synchronization and Identification Performance}

\begin{figure}[!t]
    \centering
    \subfloat[$U_b=2$.]{%
        \includegraphics[width=0.5\linewidth]{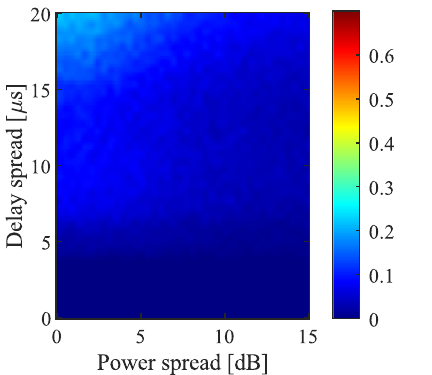}%
        \label{fig:dspread_vs_power2}}
    \hfill
    \subfloat[$U_b=4$.]{%
        \includegraphics[width=0.5\linewidth]{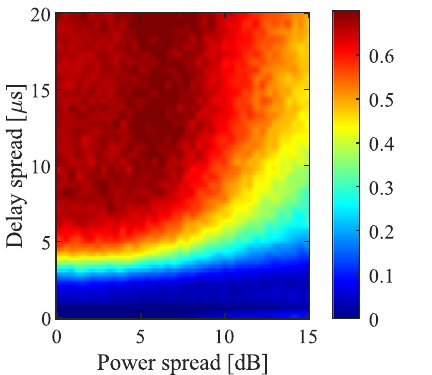}%
        \label{fig:dspread_vs_delay4}}
    \caption{Misidentification probability for (a) $U_b = 2$ and (b) $U_b = 4$ as a function of the inter-\gls{ue} received power disparity and delay spreads, for an SRS bandwidth of 1.4\,MHz.}
    
    \label{fig:dspread_vs_power}
\end{figure}
\begin{figure}[!t]
    \centering
    \subfloat[]{%
        \includegraphics[width=0.47\linewidth]{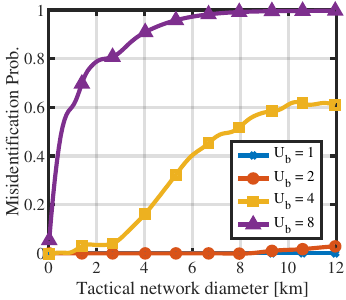}%
        \label{fig:dspread_low_B}}
    \hfill
    \subfloat[]{%
        \includegraphics[width=0.47\linewidth]{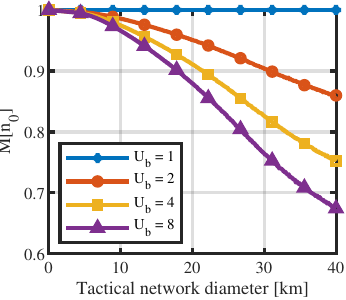}%
        \label{fig:M_vs_diameter}}
 \caption{Misidentification probability (a) and synchronization metric $M[n_0]$  (b) as a function of the network diameter  for different numbers of UEs sharing the same SRS time-frequency resources ($U_b = 1, 2, 4, 8$). Increasing the diameter leads to larger delay spreads, which degrade both synchronization performance and \gls{ue} identification reliability.}
    \label{fig:misid_bandwidth}
\end{figure}

We first evaluate the robustness of synchronization and the reliability of multi-\gls{ue} identification as the deployment scale increases. In particular, we analyze how the synchronization metric degrades with network diameter, and how the misidentification probability depends on the relative delay and power spreads among UEs sharing the same SRS subband $\mathcal{K}_b$.

To this end, UEs are randomly distributed over scenarios with increasing diameter, assuming equal transmit power. This represents a worst-case condition, as in practical deployments \glspl{ue} are often spatially clustered, leading to smaller delay spreads for a given area.

Fig.~\ref{fig:dspread_vs_power} shows the misidentification probability as a function of delay and power spreads for different numbers of multiplexed \glspl{ue}. When $U_b = 2$, identification remains highly reliable even for large delay spreads. As the deployment diameter increases, the delay spread grows, but this is typically accompanied by a corresponding increase in power disparity due to path loss. As a result, one \gls{ue} tends to dominate the received signal, and the misidentification probability remains below 1\% even for large-scale scenarios (up to $\sim 10$\,km).

For $U_b = 4$, the impact of delay spread becomes more pronounced. In this case, reliable identification (below 1\% misidentification probability) is maintained only for smaller deployment areas (on the order of $\sim 1$\,km), highlighting a trade-off between multiplexing capability and deployment scale. These trends are further summarized in Fig.~\ref{fig:dspread_low_B}, which reports the misidentification probability versus network diameter. Larger values of $U_b$ (e.g., $U_b = 8$) are suitable only for small-scale deployments, where delay spreads remain limited.

From a practical perspective, for mission scenarios involving lightweight non-serving UAVs operating over predefined areas of interest (typically on the order of hundreds of meters), the derived bounds are sufficient to ensure reliable, and in practice near error-free, instantaneous \gls{ue} identification even under $U_b$$\,=\,$$4$ multiplexing conditions. 
This enables a single lightweight \gls{uav} to simultaneously collect measurements from all \glspl{ue} within the same subband, drastically reducing complexity and mission times.

It is important to note that the reported misidentification probabilities refer to correct identification at the initial stage, i.e., when the UAV is still far from the \glspl{ue}. In practice, even in cases where identification is uncertain at large distances, the system can reliably recover the correct \gls{ue} identity during the refinement phase. As the UAV approaches a given UE, its contribution becomes dominant, effectively reducing both delay spread and interference from other \glspl{ue}, and enabling accurate re-identification. This is particularly relevant when a specific UE must be identified from the outset, whereas in search scenarios the system naturally converges to all \glspl{ue} during the refinement phase.

In contrast, synchronization remains robust across all considered scenarios. As shown in Fig.~\ref{fig:M_vs_diameter}, the synchronization metric $M[n_0]$ remains well above the detection threshold even for large deployment diameters (up to $40$\,km) and highest multiplexing levels ($U_b$$\,=\,$$8$), indicating that time alignment is not a limiting factor even under significant delay spreads.

Finally, it is worth noting that the presented results correspond to a narrowband configuration (1.4\,MHz). Increasing the SRS bandwidth would reduce cyclic shift inter-bin leakage (see Fig.~\ref{fig: C[k]}), thereby enhancing identification performance. This comes at the cost of increased hardware complexity and reduced multiplexing capability across SRS subbands, due to the additional bandwidth required per subband.

\begin{figure}[!t]
    \centering
    \subfloat[]{%
    \includegraphics[width=0.24\textwidth]{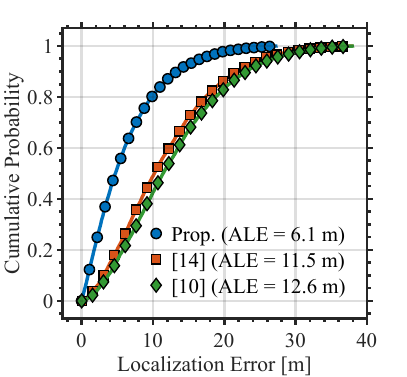}%
        }
           \hfill
    \subfloat[]{%
       \includegraphics[width=0.24\textwidth]{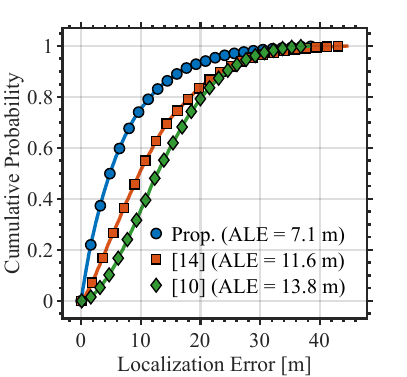}%
        }
  \caption{CDFs of the localization error obtained with the proposed method and benchmark approaches in (a) rural and (b) urban environments.}
    \label{fig:cdf}
\end{figure}

\begin{figure*}[!ht]
    \centering
    \subfloat[Hardware components and connections.]{
\includegraphics[width=0.33\linewidth]{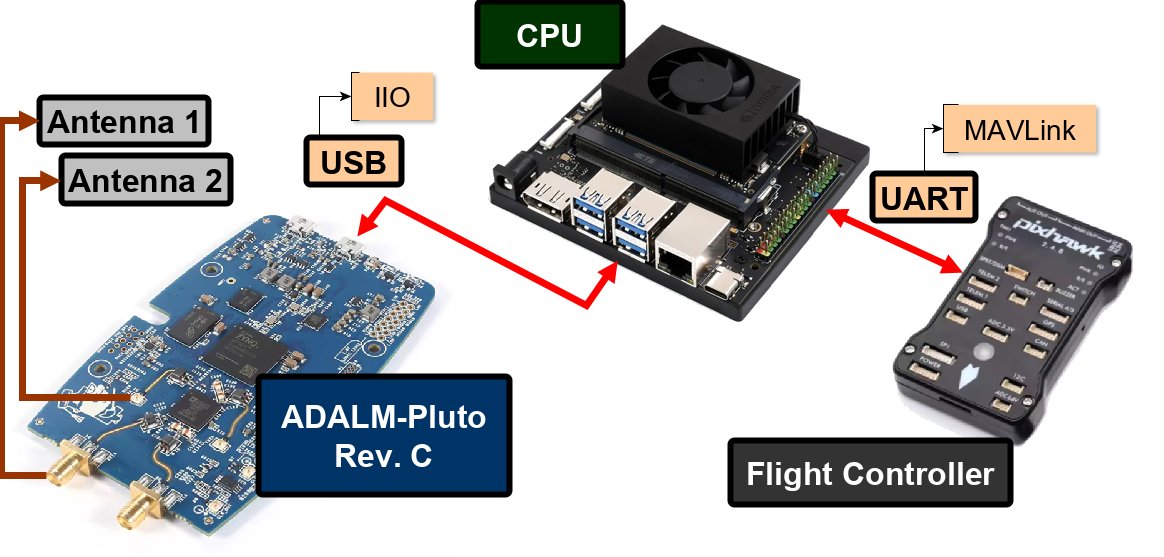}
        \label{fig:JetsonPixhawk}
    }
    \subfloat[Assembled F450 UAV.]{
  \includegraphics[width=0.33\linewidth]{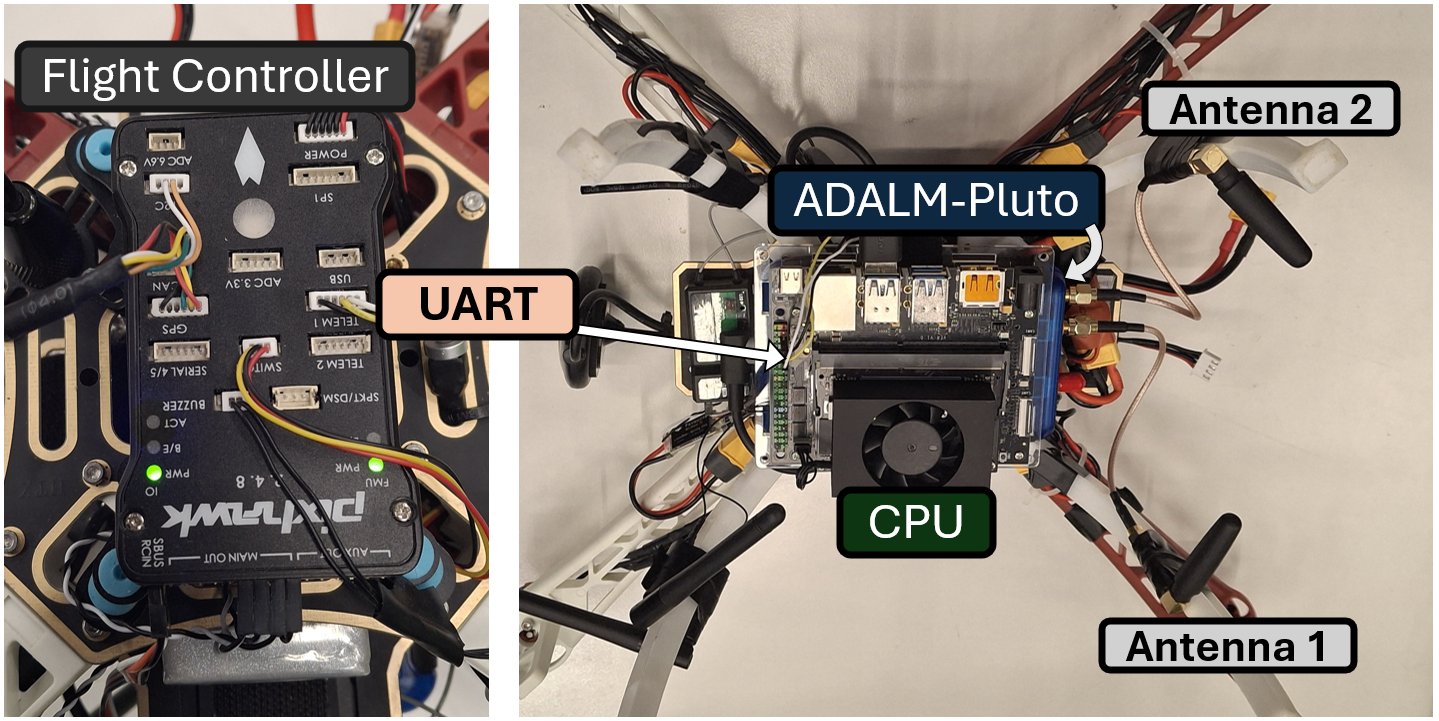}
        \label{fig:photochild}
    }
    \subfloat[On field deployment.]{
  \includegraphics[width=0.3\linewidth]{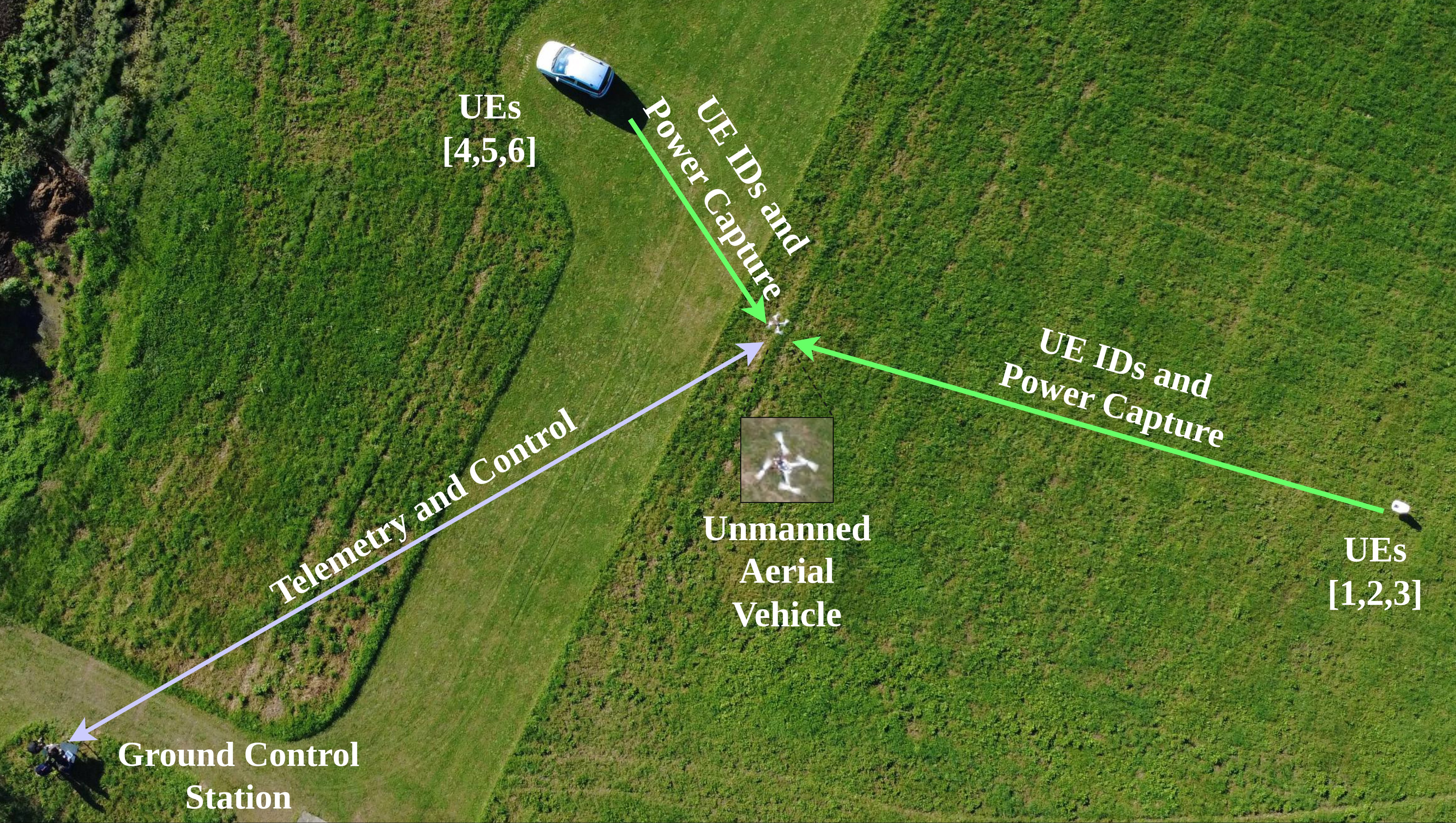}
        \label{fig:testbed}
    }
    \caption{Hardware, physical realization, and experimental testbed of the autonomous UAV-based localization platform. 
    (a) Onboard hardware architecture. 
    (b) Assembled F450 drone with payload and antenna system. 
    (c) On field deployment and experimental setup for UAV-based localization of six UEs.}
    \label{fig:drone_system_full}
\end{figure*}

\subsection{UEs Localization Performance}

We compared the proposed localization approach against two experimental state-of-the-art methods \cite{scazzoli2023experimental, esrafilian2025first}, while using the same synchronization and identification procedures (Sec.~\ref{sec:workflow}) across all implementations to ensure consistent measurement extraction and user identification.

The first benchmark follows a \gls{tdoa}-based scheme that exploits the \gls{uav} motion around the targets, whereas the second performs triangulation using the \gls{aoa} estimated at each \gls{uav} position. 

The proposed framework employs individual narrowband \gls{srs} allocations of $1.4\,\mathrm{MHz}$ (4~\glspl{rb}, \gls{5g} numerology~1), up to $20\,\mathrm{MHz}$ overall when multiple subbands are used. This choice reduces UE power consumption, lowers hardware and processing complexity, and increases multiplexing flexibility, as multiple \glspl{ue} can share different subbands or cyclic shifts within the same SRS subband. In each simulation, four \glspl{ue} were deployed per SRS band, enabling simultaneous measurement collection, coarse localization, and reduced mission times.

Unlike conventional approaches that rely on resolving individual multipath components (e.g., AoA or TDoA estimation), the proposed method leverages the spatial evolution of the SRS-derived metric $\gamma[r]$ for each user along the UAV trajectory. This trajectory-based processing captures large-scale channel variations, while the dense set of measurements collected along the flight path naturally mitigates small-scale fading effects.

For a fair comparison, the benchmark methods operated over a $40\,\mathrm{MHz}$ single SRS subband to mitigate multipath effects and improve delay estimation of the dominant \gls{los} component, which is essential for accurate \gls{toa}- and \gls{aoa}-based approaches. However, the wider bandwidth entails higher sampling rates, greater receiver complexity, and stricter synchronization requirements. Moreover, maintaining the same \gls{snr} over a larger bandwidth requires higher UE transmit power, increasing energy consumption and reducing multiplexing capability, as each user occupies the full SRS bandwidth.

To evaluate the positioning method, we use the Localization Error (LE) metric, defined for the \( u \)th \gls{ue} as $\mathrm{LE}_{\!u} = \sqrt{ \left\| \hat{\mathbf{m}}_u - \mathbf{m}_u \right\|^2 }.$ 

The Average Localization Error (ALE) is then computed as the mean LE over all localized \glspl{ue}. For each of the two considered scenarios, 20 independent statistical channel realizations were generated, with 20 UAV flights simulated for each realization. The cumulative distribution functions (CDFs) of the positioning error obtained with the three methods, along with their corresponding ALE values, are shown in Figs.~\ref{fig:cdf}a and \ref{fig:cdf}b. In the rural scenario, the proposed method achieves an ALE of 6.1\,m, compared to 11.5\,m and 12.6\,m for \cite{scazzoli2023experimental, esrafilian2025first}, with corresponding standard deviations of 5.1\,m, 7.1\,m, and 7.4\,m, respectively. In the urban scenario, the corresponding ALE values are 7.1\,m, 11.6\,m, and 13.8\,m, with standard deviations of 7.4\,m, 8.7\,m, and 7.5\,m, respectively. Overall, the proposed approach achieves an average localization improvement of approximately 5--6\,m compared to \cite{scazzoli2023experimental, esrafilian2025first}.

At low flight altitudes
\gls{aoa} tends to approach the horizon, which limits its localization capability.
In real deployments, it also requires multi-element arrays with precise geometry and onboard 
calibration of both the \gls{rf} chains and antennas, which is extremely difficult to support on small aerial platforms. 
\gls{tdoa} based localization faces similar constraints: accurate multilateration needs sub-nanosecond synchronization, and even small timing offsets create meter-level errors.
In a passive sensing scenario such as ours, the \gls{ue} clock bias cannot be eliminated through two-way techniques (e.g., Round Trip Time (RTT)), since the UAV operates as a receiver only. 
As a result, TDoA inherently depends on the UE clock stability and on how frequently network-side timing corrections are applied, parameters that are not under the control of either the network or the sensing platform and introduce additional drift-induced errors \cite{JSAC_Paglierani}.

\section{Experimental Campaign} \label{sec:experiments}
The \gls{uav} control and telemetry architecture used for experimental validation of the proposed \glspl{ue} identification and localization system is described below.

\subsection{Testbed Configuration}
\label{subsec:testbed config}

The experimental setup consists of an autonomous UAV platform and a GCS.  The UAV communicates with the GCS via Wi-Fi, transmitting telemetry data (position, velocity, system status) over a UDP link, which are visualized in real time through a custom interface. 

In this architecture, autonomy refers to the UAV’s ability to process uplink signals, identify \glspl{ue}, estimate their positions, and update its flight trajectory in real time without external control or offboard computation.

The onboard system integrates three main components: an SDR, a processing unit, and a flight controller, as illustrated in Fig.~\ref{fig:JetsonPixhawk}. The RF front-end is based on an ADALM-Pluto SDR (Rev. C), modified to enable dual-channel reception. Two stock dipole antennas, mounted approximately $30\,\text{cm}$ apart and oriented orthogonally, provide spatial diversity and wide angular coverage. The SDR streams I/Q samples to the onboard processor via USB.
The processing unit is an NVIDIA Jetson Orin NX, which handles signal processing, UE identification, localization, and mission control. The flight controller is a Pixhawk 2.4.8 running ArduPilot, enabling autonomous waypoint navigation and real-time telemetry via MAVLink. Position estimates are provided by an M8N Global Navigation Satellite System (GNSS) receiver with $0.5\,$m nominal accuracy. The complete UAV platform is shown in Fig.~\ref{fig:photochild}.

The experimental campaign was conducted at the HeliFly field near Milan, in an open-area environment suitable for UAV-based sensing (see Fig.~\ref{fig:testbed}). Six UEs were deployed: three placed outdoors at approximately $1\,\text{m}$ height and three inside a vehicle. The ground-truth positions of all UEs were collected using the GNSS receiver of a commercial smartphone.

UEs transmitted standard 5G uplink signals with SRS enabled and configured using the proposed optimized parameters prior to the mission. Two SRS subbands were used, each supporting three UEs via cyclic shifts. The UAV monitored both subbands simultaneously, treating all \glspl{ue} as independent targets.
The carrier frequency was set to $866\,\text{MHz}$, with UE transmit power below $10\,\text{dBm}$ and UAV receiver gain of $50\,\text{dB}$. During flight, the UAV collected \gls{srs} measurements and corresponding positions for localization.
\begin{figure}[!t]
    \centering
    \subfloat[]{%
        \includegraphics[width=\linewidth]{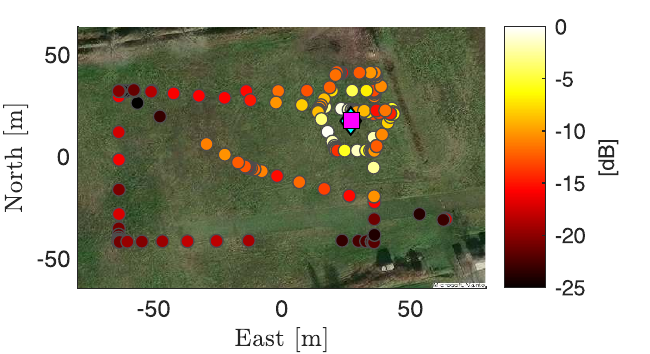}%
    }
    
    \subfloat[]{%
        \includegraphics[width=\linewidth]{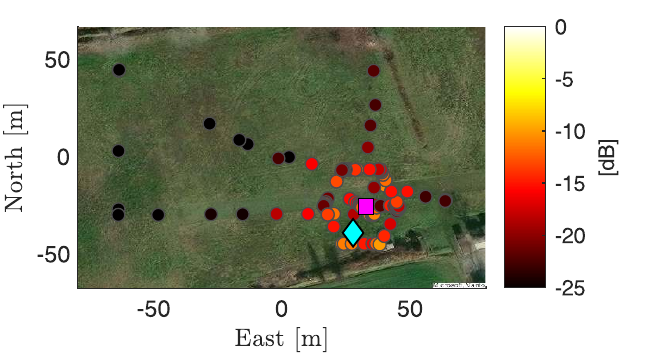}%
    }
   \caption{Measured $\tilde {\boldsymbol{\gamma}}^{(u)}_{SRS}[r]$ and localization results obtained during the experimental campaign: 
(a) \gls{ue}~1 with clear \gls{los} condition, 
(b) \gls{ue}~4 placed inside a vehicle under NLoS conditions. 
 Markers indicate the estimated (purple) and true (light blue) \gls{ue} positions.}
    \label{fig:powermap}
\end{figure}
\begin{table}[!t]
\captionsetup{justification=raggedright,singlelinecheck=false}
\fontsize{7}{8}\selectfont
\caption{Average experimental results.}
\label{tab:exp_results}
\renewcommand{\arraystretch}{0.8}
\setlength{\tabcolsep}{3pt} 
\centering
\scriptsize
\rowcolors{2}{gray!15}{white} 
\begin{tabular}{p{0.1\linewidth} p{0.18\linewidth} p{0.18\linewidth} p{0.18\linewidth} p{0.18\linewidth}}
\toprule
\textbf{UE ID} & \multicolumn{2}{c}{\textbf{Flight Duration [min]}} & \multicolumn{2}{c}{\textbf{Localization Accuracy [m]}} \\
\cmidrule(lr){2-3} \cmidrule(lr){4-5}
 & Initial & Refined & Initial & Refined \\
\midrule
UE1 & 1.55 & 3.35 & 2.8 ± 1.6 & 1.8 ± 0.6 \\
UE2 & 1.55 & 4.55 & 2.8 ± 1.6 & 1.8 ± 0.6 \\
UE3 & 1.55 & 6.15 & 2.8 ± 1.6 & 1.7 ± 0.5 \\
UE4 & 1.55 & 8.05 & 10.1 ± 1.5 & 8.2 ± 1.2 \\
UE5 & 1.55 & 8.35 & 10.3 ± 1.3 & 8.3 ± 1.1 \\
UE6 & 1.55 & 10.05 & 9.9 ± 1.4 & 8 ± 1.0 \\
\bottomrule
\end{tabular}
\end{table}

\subsection{Experimental Results}

The UAV autonomously executed the sensing mission, collecting measurements over the area of interest and performing both initial localization and refinement for all UEs.
Representative results are shown in Fig.~\ref{fig:powermap}. For outdoor UEs under LoS conditions, measurements were available across most of the area, enabling localization errors below $4.5\,\text{m}$ after the initial estimate and below $2.5\,\text{m}$ after refinement. 

For UEs placed inside the vehicle, propagation was affected by blockage from the metallic structure, resulting in lower received power and less regular spatial patterns. In this case, localization errors remained below $10\,\text{m}$ after both estimation stages. As expected, the received power maxima were not aligned with the UE position but were influenced by reflections and partial visibility through windows.

Table~\ref{tab:exp_results} summarizes the results obtained from 22 experimental flights. On average, the initial position estimates for all UEs were obtained in approximately $2.15\,\text{min}$, followed by an additional $1.30\,\text{min}$ per UE for the refinement phase.

Measurement density was limited by the SDR hardware (USB 2.0), resulting in an effective sampling interval of approximately $1\,\text{s}$. This prevented continuous SRS tracking along the trajectory and led to a relatively sparse spatial sampling. We therefore expect that denser measurements, achievable with higher-performance SDR platforms, would further improve localization accuracy.

Reliable coarse synchronization was observed across the measurement area when UEs operated under sufficient SNR conditions (see Fig.~\ref{fig:powermap}a). This behavior is consistent with the analysis in Sec.~\ref{sec:Coarse time sync} and the simulation results in Fig.~\ref{fig:M_vs_diameter}, which show that the synchronization metric remains stable even under large delay spreads.

Multi-\gls{ue} identification was correctly performed in all experimental runs, with no misidentification events observed for the considered configurations. This agrees with the analysis in Sec.~\ref{sec:Coarse time sync}, which indicates that for a limited number of \glspl{ue} (here $U_b $$\,=\,$$ 3$) and deployment radii below $\sim 1$\,km, the probability of misidentification remains negligible. The results also demonstrate that all operations were correctly performed with three multiplexed UEs sharing the same time--frequency SRS resources. This enables simultaneous measurement collection for all \glspl{ue}, significantly reducing the overall mission time while maintaining reliable identification and localization performance on a low-complexity hardware platform.

Overall, the experimental results agree with the theoretical and simulation analysis, showing that the proposed framework operates under realistic UAV-based deployment conditions.

\section{Conclusion and future works}\label{sec:conclusion}

This work demonstrated that a non-serving lightweight UAV can autonomously identify and localize multiple \glspl{ue} by passively exploiting standard 5G uplink \gls{srs} transmissions, without network synchronization or real-time control-plane interaction.
The proposed framework was validated through simulations and a full-scale experimental campaign, achieving sub-8\,m accuracy in urban scenarios and sub-3\,m in rural conditions.
The results highlight that reliable synchronization and identification can be achieved even under worst-case multiplexing conditions, where multiple \glspl{ue} share the same narrowband SRS subband (1.4\,MHz). This confirms that meter-level localization is feasible with lightweight hardware, reduced bandwidth, and limited UE energy consumption, making the approach suitable for practical UAV deployments.
These findings support the use of sensing-oriented \glspl{uav} as a complementary solution for infrastructure-independent situational awareness in \gls{lawn} scenarios.
Future work will focus on cooperative multi-UAV operation to improve scalability and reduce mission time, leveraging coordinated trajectories and shared measurements among multiple platforms to enhance robustness and awareness in complex or large-scale environments. Additionally, extending the framework to support highly dynamic targets through adaptive trajectory replanning will be investigated, building upon the current design.

\appendices
\section{Expansion of the Detection Metric}
\label{app: expansion of detection metric}
Using the notation introduced in Section~\ref{sec:Coarse time sync}, the detection metric, when an SRS is received, evaluated at the  reference arrival point $n_0$ can be expressed as
\begin{equation}
    M[n_0] = \frac{|P[n_0]|^2}{|R[n_0]|^2}.
\end{equation}

The correlation term $P[n_0]$, neglecting the noise contribution, was first expanded exactly as
\begin{align}
& P[n_0] = \sum_{l=0}^{L-1}
\left(
   \sum_{u=1}^{U}\sum_{p=1}^{P_u} 
   a_{p,u}\, x_u(lT_s - \Delta \tau_{p,u})
   e^{j2\pi \nu_{u,\mathrm{cfo}} lT_s}
\right) \notag \\[6pt]
&\times
\left(
   \sum_{v=1}^{U}\sum_{q=1}^{P_v} 
   a_{q,v}^*\, x_u^*((l+L)T_s - \Delta \tau_{q,v})
   e^{-j2\pi \nu_{v,\mathrm{cfo}} (l+L)T_s}
\right),
\end{align}
where we assumed $\nu_{u,\mathrm{cfo}} \gg \nu_{p,u}$.

\subsubsection{Single-Path, Single-User Contribution}

For the case \(u=v\) and \(p=q\), the correlation term simplifies, up to a common phase rotation, to
\begin{small}
\begin{equation}
\label{eq:P_n0_expanded}
P[n_0] =
\sum_{u=1}^{U}\sum_{p=1}^{P_u} |a_{p,u}|^2
\sum_{l=0}^{L-1}
x_{u}\!\big(lT_s-\Delta\tau_{p,u}\big)\,
x_{u}^*\!\big((l+L)T_s-\Delta\tau_{p,u}\big).
\end{equation}
\end{small}
Assuming normalized symbol energy \(E_s=1\), and no additional \gls{ofdm} symbols before or after the target \gls{srs}, we define the normalized correlation of a single path as
\begin{equation}
f_0(\Delta\tau_{p,u})
=
\sum_{l=0}^{L-1}
x_{u}\!\big(lT_s-\Delta\tau_{p,u}\big)\,
x_{u}^*\!\big((l+L)T_s-\Delta\tau_{p,u}\big),
\end{equation}
such that
\begin{equation}
P[n_0]
\approx
\sum_{u=1}^{U}\sum_{p=1}^{P_u}
|a_{p,u}|^2
f_0(\Delta\tau_{p,u}).
\end{equation}
The function \(f_0(\Delta\tau_{p,u})\) depends on the timing offset since only part of the repeated SRS symbol is observed within the two correlation windows. As the timing offset varies, an increasing portion of the repeated waveform falls inside the correlation windows, increasing the correlation value. Once the timing error is completely absorbed by the cyclic prefix, the correlation reaches its maximum and remains constant. Finally, as the repeated waveform progressively leaves the correlation windows, the correlation decreases.

If the repeated SRS waveform had constant instantaneous power, this behavior would result in an exact piecewise-linear function. In practice, however, the instantaneous power of \(s_u[n]\) is not perfectly constant because only \(M_{\mathrm{SRS}}\) active subcarriers are mapped over an IDFT of size \(N\). Therefore, \(f_0(\Delta\tau_{p,u})\) is  well approximated by

\begin{small}
\begin{equation}
\label{eq:f_tau}
f_0(\Delta\tau_{p,u}) \approx
\begin{cases}
\dfrac{\Delta\tau_{p,u}+T_{CP}+T_L}{T_L},
&
\hspace{-0.3cm}-T_L-T_{CP}
\le
\Delta\tau_{p,u}
<
-T_{CP},
\\[5pt]
1,
&
-T_{CP}
\le
\Delta\tau_{p,u}
<
0,
\\[5pt]
1-\dfrac{\Delta\tau_{p,u}}{T_L},
&
0
\le
\Delta\tau_{p,u}
<
T_L,
\\[5pt]
0,
&
\text{otherwise},
\end{cases}
\end{equation}
\end{small}

where \(T_L=LT_s\). No residual \gls{cfo} appears in \(f_0(\Delta\tau_{p,u})\), since the phase rotations affecting the two repeated SRS halves cancel out.

\subsubsection{Multi-Path Contribution}
When $u=v$ and $p \neq q$, the expression (up to a common phase term) becomes 
\begin{align}
P[n_0] &= \sum_{u=1}^{U} \sum_{\substack{p,q \\ p \neq q}} 
   a_{p,u} a_{q,u}^* 
   \sum_{l=0}^{L-1} 
   x_{u}(lT_s - \Delta \tau_{p,u}) \notag \\[6pt]
&\quad \times x_{u}^*((l+L)T_s - \Delta \tau_{q,u}) \,
   e^{j2\pi (\nu_{p,u}-\nu_{q,u})lT_s}.
\end{align}
Due to the impulsive autocorrelation property of \glspl{srs}, delayed replicas of the same signal are nearly orthogonal, and the above term vanishes, i.e., $P[n_0] \approx 0$.

\subsubsection{Multi-User Contribution}
For the case $u \neq v$, we obtain
\begin{align}
P[n_0] &= \sum_{\substack{u,v \\ u \neq v}} \sum_{\substack{p,q}}
     a_{p,u}a_{q,v}^*  \sum_{l=0}^{L-1}
     x_u(lT_s - \Delta \tau_{p,u})\,\notag \\[6pt]
&\quad \times 
     x_v^*((l+L)T_s - \Delta \tau_{q,v})
     e^{j2\pi \Delta\nu lT_s}.
\end{align}
where $\Delta\nu = \nu_{u,\mathrm{cfo}} - \nu_{v,\mathrm{cfo}}$.  
This term can be neglected under standard cellular conditions, since \glspl{srs} associated with different cyclic shifts are quasi-orthogonal and \gls{cfo} values are typically limited to a few hundred hertz. The approximation holds when 
\begin{equation}
\label{eq:orthogonality condition}
 \Delta\tau_{p,u} - \Delta\tau_{q,v} \neq (\eta_u - \eta_v)T_sL/8,   
\end{equation}

 that is, when the relative propagation delay between two received SRS signals does not coincide with the time-domain cyclic shift separating the corresponding transmitted SRS sequences.

\begin{figure}[!t]
    \centering
    \includegraphics[width=0.9\linewidth]{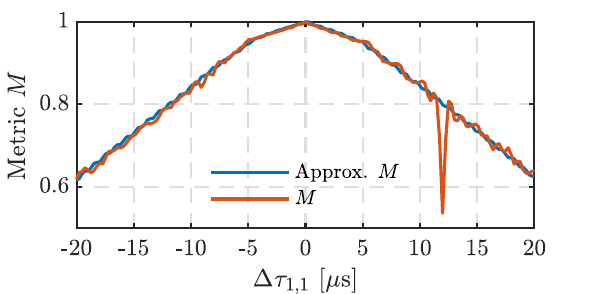}
    \caption{Comparison between the exact correlation metric $M[n]$ (red), and its approximation (blue).}
    \label{fig:M_real_approx}
\end{figure}
The normalization term $R[n_0]$ is given by
\begin{equation}
R[n_0] =
\tfrac{1}{2}
\left(
\sum_{l=0}^{L-1}|y[l]|^2
+
\sum_{l=0}^{L-1}|y[l+L]|^2
\right).
\end{equation}
Expanding the squared magnitude produces diagonal terms $(u=v,p=q)$ and off-diagonal terms corresponding to different propagation paths or different UEs. Exploiting the same orthogonality properties discussed above, the off-diagonal terms become negligible, and the dominant contribution becomes
\begin{align}
R[n_0]
&\approx
\sum_{u=1}^{U}\sum_{p=1}^{P_u}
|a_{p,u}|^2 g_0(\Delta\tau_{p,u}),
\end{align}
where
\begin{align}
g_0(\Delta\tau_{p,u})
&=
\tfrac{1}{2}
\sum_{l=0}^{L-1}
\Big[
|x_u(lT_s-\Delta\tau_{p,u})|^2
\notag\\
&\qquad\qquad+
|x_u((l+L)T_s-\Delta\tau_{p,u})|^2
\Big].
\end{align}

Assuming normalized symbol energy $E_s=1$, $g_0(\Delta\tau_{p,u})$ can be approximated in piecewise form as
\begin{small}
\begin{equation}
\label{eq:g_tau}
g_0(\Delta\tau_{p,u}) \approx
\begin{cases}
  \dfrac{\Delta\tau_{p,u} \hspace{-0.05cm}+ \hspace{-0.05cm}T_{CP} \hspace{-0.05cm}+ \hspace{-0.05cm}2T_L}{2T_L}, 
  & \hspace{-0.2cm}-\hspace{-0.05cm}2T_L \hspace{-0.05cm}- \hspace{-0.05cm}T_{CP} 
    \hspace{-0.05cm}\leq \hspace{-0.05cm}\Delta\tau_{p,u} 
    \hspace{-0.05cm}< \hspace{-0.05cm}-\hspace{-0.05cm}T_{CP}, \\[4pt]
  1, 
  & -T_{CP} \leq \Delta\tau_{p,u} < 0, \\[4pt]
  1 - \dfrac{\Delta\tau_{p,u}}{2T_L}, 
  & 0 \leq \Delta\tau_{p,u} < 2T_L, \\[4pt]
  0, 
  & \text{otherwise}.
\end{cases}
\end{equation}
\end{small}

Figure~\ref{fig:M_real_approx} shows the real metric compared to the approximated one. Around $\Delta\tau_{1,1} = 13\,\mu$s, a notch appears due to the violation of the orthogonality condition in~\eqref{eq:orthogonality condition}.



\ifCLASSOPTIONcaptionsoff
  \newpage
\fi
\bibliographystyle{IEEEtran}
\bibliography{references}

\end{document}